\documentclass[a4paper,12pt]{article}
\pdfoutput=1
\usepackage{amsmath}
\usepackage{amssymb}
\usepackage{amsthm}
\usepackage{graphicx}
\usepackage{caption}
\usepackage{subfig}
\usepackage{color}

\usepackage[utf8]{inputenc}

\usepackage[top=3cm, bottom=5cm, left=2cm, right=2cm]{geometry}

\newtheorem{thm}{Theorem}
\newtheorem{lem}{Lemma}[thm]

\theoremstyle{remark}
\newtheorem{rem}{Remark}[section]

\theoremstyle{definition}

\newcounter{todo}[section]

\newenvironment{example}{{\bf Example : }}{$\hfill\square$}

\newcommand{\dfp}[2]{\partial_{#1} f^\prime_{#2}}

\def\irdeu{\frac{1}{\sqrt{2}}}

\def\M{\mathcal{M}}

\def\G{\Gamma}
\def\MG{\mathcal{M}_\Gamma}
\def\Mp{\mathcal{M}_2}

\def\Mall{\mathcal{M}_{all}}

\def\g{\gamma}
\def\Mg{\mathcal{M_\gamma}}
\def\Vg{V_\gamma}

\def\sp{s^\prime}
\def\up{u^\prime}

\def\NatO{N_{\alpha,\|.\|}^{t_0,t}}
\def\Na{N_{\alpha,\|.\|}^{-\infty,t}}

\def\Nb{N_{\beta,\|.\|}^{-\infty,t}}

\def\Vp{V_{all}^\prime}
\def\Vpt{V_{all}^{\prime T}}

\def\fp{f^\prime}
\def\np{n^\prime}
\def\wp{w^\prime}

\author{Léonard Gérard and Jean-Jacques Slotine}
\title{Neuronal networks and controlled symmetries, \\ a generic framework}
\date{}

\begin{document}

\maketitle

\begin{abstract}
The extraordinary computational power of the brain may be related in
part to the fact that each of the smaller neural networks that compose
it can behave transiently in many different ways, depending on its
inputs. We use contraction theory to extend earlier work on synchrony
and symmetry, and exploit input continuity of contracting systems to
ensure robust control of spatial and spatio-temporal symmetries of the
output with the input.
\end{abstract}

\section{Introduction}
The brain is often described as being composed, in part, of a very large
number of small ``identical'' functional
units~\cite{Mountcastle1978}. Cortical computation, for instance, is
commonly viewed as being organized around cortical
columns~\cite{Koerner1999,Kupper2006,Milo2002,Hayon2005}. In this
context, a frequent suggestion is that the overall computational power
of the brain may be related to some sort of combinatorial
complexity~\cite{Malsburg1995}, and to the fact that each part of the
brain is reused for different computations.

At the level of individual units, although high behavioral variety
could be explained by some learning process ~\cite{Robert1993,Bienenstock1982,Abarbanel2002,Malenka2004,Izhikevich2004}
or internal change,it is unlikely that such changes occur in very short
periods of time. Instead, the most efficient source of behavioral variety
could be simply to have a high input dependency, exploiting the
nonlinearity of biological neural networks. In other words, depending
on its input, a functional unit could behave in very diverse ways,
though remaining stable and robust against noise and small variations
in the input.

Intuitively, it is also known that sensory ``input'' enrolls only 5\%
or so of the connections to the thalamus~\cite{Sherman2002}, and that
a similarly small percentage describes connections from the thalamus
to input cortical layers~\cite{Binzegger2004}. Recent research
suggests that this ``paucity of input'' between different regions of
the brain is actually quite general~\cite{Kennedy2007}.

Inspired by this, we will try to draw a generic framework allowing to
observe neural systems under some ``controlling'' input. To this
effect, we introduce ``input
continuity'' analysis which will provide a way to describe the
properties of a unit's output knowing the properties of its
input. This framework has broader possible uses and applications than
used in this paper and will be discussed in the appendix. Here the
main use of the ``input continuity`` will be its link with contracting
systems~\cite{Lohmiller1998}, giving us a simple way to change the
behavior of systems as will be shown with some toy examples in Section
\ref{sec:control} also displaying some interesting results about
symmetries and contracting systems, results exposed in Section
\ref{sec:symmetries} of the paper.

Indeed the study of symmetries is important in dynamical systems
\cite{Spong2005,Golubitsky1999} and more specifically in neural
networks. It strongly influences, as we will see, synchrony and
polysynchrony (concurrent synchronization), concepts playing an
important role in neurobiology
~\cite{Boergers2003,Pinsky1995,Kazanovich2003}. To this matter we will
try to cover Lie continuous symmetries and spatio-temporal symmetries,
giving some interesting tools to ensure this symmetries in the
output. Most of those are based on contracting systems. Let us first
recall different contraction theorem and properties.

\section{Contraction}
Essentially, a nonlinear time-varying dynamic system will be called {\it contracting} if initial conditions or temporary disturbances are forgotten exponentially fast, i.e., if trajectories of the perturbed system return to their nominal behavior with an exponential convergence rate.  It turns out that relatively simple algebraic conditions can be given for this stability-like property to be verified, and that this property is preserved through basic system combinations.

A nonlinear contracting system has the following properties~\cite{Lohmiller1998,Lohmiller2000,Slotine2002,Wang2004}
\begin{itemize}
	\item global exponential convergence and stability are guaranteed
	\item convergence rates can be explicitly computed as eigenvalues of well-defined symmetric matrices
	\item robustness to variations in dynamics can be easily quantified
\end{itemize}

\subsection{Basic results}
Our general dynamical systems will be in $\mathbb{R}^n$, deterministic, with $f$ a smooth non linear function.
\begin{equation}
\label{eq:main}
	\dot x = f(x,t)
\end{equation}

The basic theorem of contraction analysis, derived in~\cite{Lohmiller1998}, can be stated as:
\begin{thm}[Contraction]
	Denote the Jacobian matrix of $f$ with respect to its first variable by $\frac{\partial f} {\partial x}$. If there exists a square matrix $\Theta(x,t)$ such that $\Theta(x,t)^T\Theta(x,t)$ is uniformly positive definite and the matrix
	\[
	F = \left(\dot\Theta + \Theta \frac{\partial f} {\partial x}
	\right) \Theta^{-1}
	\]
	is uniformly negative definite, then all system trajectories converge exponentially to a single trajectory, with convergence rate $|\sup_{x,t}\lambda_\mathrm{max}(F)|>0$. The system is said to be \emph{contracting}, $F$ is called its \emph{generalized Jacobian}, and $\theta(x,t)^T\Theta(x,t)$ its contraction \emph{metric}.
\end{thm}

It can be shown conversely that the existence of a uniformly positive definite metric \[M(x,t)=\Theta(x,t)^T\Theta(x,t)\] with respect to which the system is contracting is also a necessary condition for global exponential convergence of trajectories~\cite{Lohmiller1998}.  Furthermore, all transformations $\Theta$ corresponding to the same $M$ lead to the same eigenvalues for the symmetric part $F_s $ of $F$~\cite{Slotine2002}, and thus to the same contraction rate $|\sup_{x,t}\lambda_\mathrm{max}(F_s)|$.

\begin{rem}
	In the linear time-invariant case, a system is globally contracting if and only if it is strictly stable, and $F$ can be chosen as a normal Jordan form of the system with $\Theta$ the coordinate transformation to that form~\cite{Lohmiller1998}.
\end{rem}

\begin{rem}
	Contraction analysis can also be derived for discrete-time systems and for classes of hybrid systems~\cite{Lohmiller2000}.
\end{rem}

Finally, it can be shown that contraction is preserved through basic system combinations, such as parallel combinations, hierarchies, and certain types of negative feedback, see~\cite{Lohmiller1998} for details.


\subsection{Contraction toward a linear subspace}
The main theorem of~\cite{Pham2006} gives us the ability to prove contraction of all solutions to a subspace $\M$ of the state space. It is a powerful tool that we will use in the symmetry studies, and can be stated as
\begin{thm}
	\label{thm:cuong_convergence}
	Consider a linear flow-invariant subspace $\M$ of the system ($f(\M)\subset\M$) and the associated orthonormal projection matrix $U^TU$ (we have $V^TV + U^TU = I_n$ and $x\in \M \iff V x = 0$). All trajectories of the system converge exponentially to $\M$ if the system
	\begin{equation}
	\label{eq:cuong_sysy}
	\dot{y}=V f(V^T y,t)
	\end{equation}
	is contracting with respect to a constant metric. If furthermore we denote the contraction rate for (\ref{eq:cuong_sysy}) by $\lambda>0$, then the convergence to $\M$ will be exponential with rate $\lambda$.
\end{thm}
We will call the above condition, $V^TfV$ contracting for a constant metric, \emph{ contraction toward $\M$}.

\begin{rem}
	\label{rem:condMg}
	The theorem uses mainly two independent hypotheses
	\begin{itemize}
		\item Contraction condition of Equation \ref{eq:cuong_sysy} : $f$ contracts toward $\M$
		\item Invariance condition of $\M$ : $f(\M) \subset \M$
	\end{itemize}
\end{rem}

\subsection{Contraction yields robustness}
It can be shown that (see section 3.7 in \cite{Lohmiller1998} for a proof and generalization)
\begin{thm}[Contraction and robustness]
	Consider a contracting system $\dot x = f(x,t)$, with a constant metric $\Theta$ and contraction rate $\lambda$. Let $P_1(t)$ be a trajectory of the system, and let $P_2(t)$ be a trajectory of the \emph{disturbed} system
	\[
	\dot x = f(x,t) + d(x,t)
	\]
	Then the distance $R(t)$ between $P_1(t)$ and $P_2(t)$ verifies $ R(t)\leq \sup_{x,t}\|d(x,t)\| / \lambda$ after exponential transients of rate $\lambda$.
\end{thm}

\section{Symmetries and contraction}
\label{sec:symmetries}
The symmetries of a neural network, defined in a broad sense, can reflect important properties. There are many different ways to express symmetries, such as symmetry of the input, the output, or the system, all of which are usually interdependent.

\subsection{Generic $\g$ operator}
Consider a dynamical system  $\dot x = f(x,t)$. A linear operator $\g$ acting
over the state space defines two usual ``symmetries'' :
\begin{itemize}
	\item symmetry of the system state: if $x=\g x$, we will say that $x$ is
$\g$-symmetric
	\item symmetry of the dynamical system : if $\g f=f \g$, we will say
that $f$ is $\g$-equivariant \cite{Golubitsky1999}
\end{itemize}
Note that any linear operator belongs to $GL$, the general linear group, see for example \cite{Spong2005} also using linear operators as symmetries.

The following simple result shows that a contracting dynamical system ``transfers'' its symmetries to its state trajectories.

\begin{lem}
	\label{lem:cont_geq}
	If $f$ is $\g$-equivariant and contracting, all solutions converge exponentially to a unique $\g$-symmetric trajectory $x(t)$.
\end{lem}
\begin{proof}
	Since the system is contracting all solutions converge exponentially to a single solution $x(t)$. But $\g x(t)$ is also a solution :
	\[
	\frac{d}{dt}\left( \g x(t) \right) = \g \dot{x}(t) = \g f(x,t) = f(\g
x,t)
	\]
	hence $x(t) \to \g x(t)$ exponentially.
\end{proof}

\subsubsection*{A simple example: permutations}
\label{sec:discrete_operator}

Let us illustrate $\g$-symmetry in the simple discrete case of a permutation operator.

Consider $x \in E = \mathbb{R}^n$, and write it as $(x_1,x_2,\dots,x_n)$, the
action of a permutation $\g$ on $E$ is defined by $\g x =
(x_{\g(1)},\dots,x_{\g(n)})$.

Decompose $\g$ into disjoint non-trivial cycles, \[\g=\sigma_0 \circ \sigma_1
\dots \sigma_p\] and decompose the space accordingly as \[E = \mathbb{R}^n =
E_{\sigma_0} \times E_{\sigma_1} \times \dots \times E_{\sigma_p} \times
E_{I_q}\] with $E_{\sigma_i}$ the space of action of $\sigma_i$.

\begin{figure}[htb]
\fbox{	\begin{minipage}{0.65\linewidth}
		\begin{align*}
			\g &= (0 1) (2 3) (4)\\
			E &= E_{(0 1)} \times E_{(2 3)} \times\mathbb{R}\\
			x &= ( x_0, x_1, x_2, x_3, x_4 )\\
			\g x &= ( x_1, x_0, x_3, x_2, x_4 )
		\end{align*}
		We have $f$ $\g$-equivariant, indeed $\g f = f \g$ :
		\begin{align*}
			f_2(x_0, x_1, x_2, x_3, x_4) &= f_1 ( x_1, x_0, x_3,x_2, x_4 )\\
			f_1(x_0, x_1, x_2, x_3, x_4) &= f_2 ( x_1, x_0, x_3,x_2, x_4 )\\
			f_4(x_0, x_1, x_2, x_3, x_4) &= f_3 ( x_1, x_0, x_3,x_2, x_4 )\\
			f_3(x_0, x_1, x_2, x_3, x_4) &= f_4 ( x_1, x_0, x_3,x_2, x_4 )\\
			f_5(x_0, x_1, x_2, x_3, x_4) &= f_5 ( x_1, x_0, x_3,x_2, x_4 )
		\end{align*}
		And $\g$ symmetry of $x$ would be synchrony of $x_0$ with $x_1$ and $x_2$ with $x_3$ :
		\begin{equation*}
			\left.
			\begin{aligned}
				x_0 = x_1  \\
				x_2 =x_3
			\end{aligned}
			\right\}   \iff \ x = \g x
		\end{equation*}
	\end{minipage}
	\begin{minipage}{0.34\linewidth}
		\includegraphics[width=0.9\linewidth]{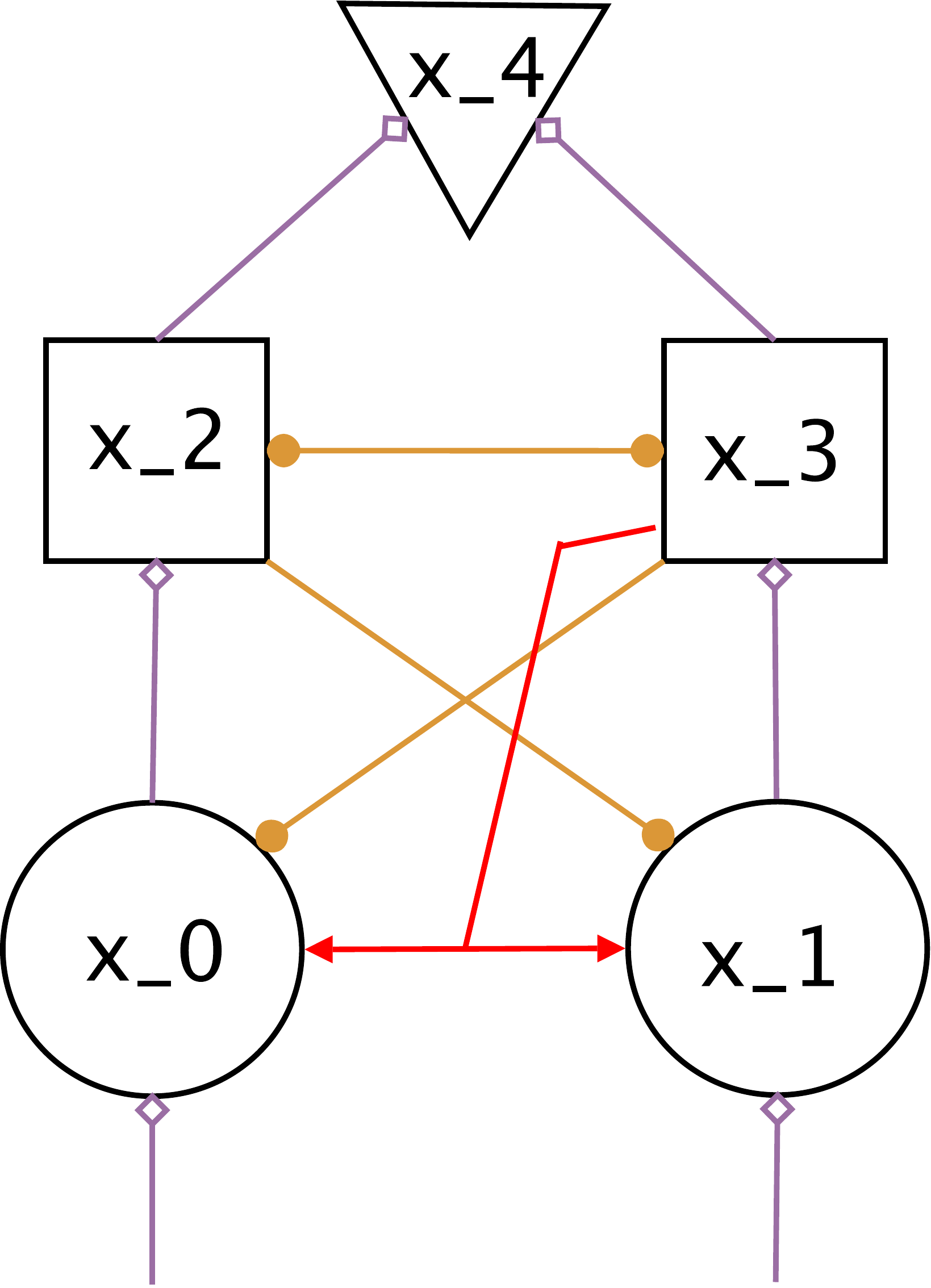}
	\end{minipage}}
	\caption{Toy example which could model a three layered network.}
	\label{fig:toy_5}
\end{figure}

$\g$ symmetry of the state space describes {\it concurrent synchronization}: in each subspace $E_{\sigma_i}$ the solution is synchronous, thus yielding $p$ co-existing synchronous assemblies, as illustrated in Figure \ref{fig:toy_5}.

\subsection {Spatio-temporal symmetries}
A straightforward extension of spatial symmetries are {\it
spatio-temporal symmetries}. Inspired by the theory developed by
Golubitsky et al. ~\cite{Golubitsky1999,Buono2001,Golubitsky2000,Josic2006}, we define a \emph{spatio-temporal symmetry $h=(\gamma,T)$}
using a spatial symmetry $\gamma$ and a period $T$, according to
\[
h x(t) = \gamma x(t+T)
\]
Unlike the $H/K$ theorem of Golubitsky et al. ~\cite{Buono2001,Josic2006}, we
do not restrict ourselves to permutation for the spatial symmetry, but, to some
linear operator $\gamma$ such as there exists an integer $p_\gamma$, classically
called the order of $\gamma$, such that $\gamma^{p_\gamma}=Id$.

\begin{example} Consider a 3-ring with the $h$-symmetry :
\[
\g \begin{pmatrix}x_1\\x_2\\x_3\end{pmatrix} =
\begin{pmatrix}\irdeu(x_1+x_3)\\x_2\\\irdeu(x_3-x_1)\end{pmatrix} \quad
\text{so} \quad \g^2 \begin{pmatrix}x_1\\x_2\\x_3\end{pmatrix} =
\begin{pmatrix}x_3\\x_2\\-x_1\end{pmatrix} \quad\text{and}\quad (\g^2)^4 = id
\]
$x$ is $h$-symmetric when :
\[
	h x(t) = x(t) \iff \left\{
	\begin{aligned}
		x_1(t) &=\irdeu \left(x_1(t+T)+x_3(t+T) \right) \\
		x_2(t) &= x_2(t+T) \\
		x_3(t) &= \irdeu \left(x_3(t+T)-x_1(t+T) \right) \\
	\end{aligned}
	\right.
\]
There is a strong interaction between $x_3$ and $x_1$ but interestingly $h x(t)
= x(t) \implies x_1(t) = x_1(t+8T) \text{ and for } x_1(0) \neq 0, x_1(t)\neq
x_1(t+T) \neq $ the same holds for $x_3$ but $x_2(t)=x_2(t+T)$. This shows that
two different rhythms have to coexist.
\end{example}

An associated definition can be given for a dynamical system. Specifically, we will say that the system $\dot x = f(x,t)$ (or its dynamics $f$) is \emph{$h$-equivariant} if
\begin{equation}
	\label{eq:hypo_h_equi}
	f(\gamma x(t),t) = \gamma f(x(t), t+T)
\end{equation}

We then obtain for spatio-temporal symmetries a result similar to Theorem~\ref{lem:cont_geq}, describing the transfer of symmetries from system dynamics to system trajectories.

\begin{thm}
\label{thm:temporalsym}
	If $f$ is contracting and $h$-equivariant, then after transients the solutions are $p_\gamma T$ periodic and exhibit the spatio-temporal symmetry $h$.
\end{thm}

Basically, all the solutions tend to a periodic solution $x_p(t)$ with the symmetry $h$ : $ x_p(t) = \gamma x_p(t+T)$. This result is an extension of the result that a periodic contracting system exhibits a unique solution of same period~\cite{Lohmiller1998}.

\begin{proof}
	If $x(t)$ is a solution, then $\gamma x(t+T)$ is also a solution:
	\[
		\frac{d}{dt}\left( \gamma x(t+T) \right) = \gamma \dot{x}(t+T) =
\gamma f(x(t+T),t+T) = f(\gamma x(t+T),t)
	\]
	Thus, since $f$ is contracting, $ x(t) \rightarrow \gamma
x(t+T)$ exponentially fast. This in turn shows that the solution tends to a
periodic signal exponentially : by recursion
	\[
	x(t) \to \g^{p_\g} x(t+p_{\g}T) = x(t+p_{\g}T)
	\]
	so that $\forall t \in  [0;p_\gamma T]$ $x(t+n p_\gamma T)$ is a Cauchy
sequence, and therefore the limiting function $\lim_{n\to \infty} x(t+n p_\gamma
T)$ exists, which completes the proof.
\end{proof}

\subsection{Spatial symmetries accomodate weaker contraction}

When dealing only with spatial symmetries, we can weaken the
contraction condition on $f$ while still transferring the symmetries
of $f$ to the system trajectory.

Conisder the linear subspace of $\g$-symmetric states, $\Mg =
\{x,\,x=\g x\}$.  Recall first a standard result linking the symmetry
of the system to this linear subspace:
\begin{lem}
	\label{lem:equi_flow}
	f is $\g$ equivariant $\implies$ $\Mg$ flow invariant
\end{lem}

\begin{proof}
	if $x \in \Mg $ and $f$ is $\g$-equivariant : $\dot{x}= f(x)=f(\g x)=\g
f(x) = \g \dot{x}$
\end{proof}

In this context, we can thus write Theorem \ref{thm:cuong_convergence} as
\begin{lem}
	\label{thm:flowcuong}
	If $\Mg$ is flow-invariant by $f$ (or sufficiently, by
Lemma~\ref{lem:equi_flow}, if $f$ is $\g$-equivariant) and $f$ contracts toward
$\Mg$, then all solutions $x(t)$ converge exponentially to $\g$-symmetric
trajectories.
\end{lem}

We will denote by $\Vg^T \Vg$ the orthogonal projector on $\Mg^\bot$ .

As these lemma shows, the more generic $\g$ is, the stronger the contraction
condition. In the generic lemma \ref{lem:cont_geq} the hypotheses of symmetry
and contraction are {\it independent} on the contrary to last lemma
\ref{thm:flowcuong} where the contraction condition depends explicitly on $\g$.

This link between symmetry and contraction is not particularly
convenient, since later in section \ref{sec:control} we will aim to
shape the equivariance of $f$, and thus the symmetries of the output,
while preserving sufficient contraction properties. We now show how to
avoid this direct dependence.

\subsection{$\Mall$ and $\MG$ spaces}
\label{sec:Mall}
We show  how to strengthen the contraction condition of lemma
\ref{lem:equi_flow} to make it independent of the symmetry condition,
or at least allowing to have the theorem hold with a set $\Gamma$ of
different symmetries.

First note:
\begin{lem}
	\label{lem:Mall2}
	The choice of $V$ to represent the orthogonal projector has no
	effect on the contraction toward $\M^\bot$.
\end{lem}

See proof in appendix \ref{proof:lemMall2}. The proof also shows that,
by contrast, using a non orthogonal projector $V_2=T V$ with $T$ square
invertible is not sufficient in general.

\begin{rem}
	\label{rem:lemMall_inv}
	The above lemma shows that the contraction condition toward a subspace is preserved when applying an orthonormal transformation $M$ to the projector $V$. Moreover one can also trivially change the metric with an orthonormal transformation : \[M \Theta V f V^T \Theta^{-1} M^T <0 \iff \Theta V f V^T \Theta^{-1} <0 \iff \Theta M V f V^T M^T \Theta^{-1} <0\]
\end{rem}

Now the main result of this section :
\begin{lem}
	\label{lem:Mall}
Consider two linear subspaces $\M$ and $\Mp$ , with $\M\subset\Mp$.
If $f$ contracts toward $\M$, then $f$ contracts toward $\Mp$.

More precisely, let $V^TV$ (resp. $V_2^TV_2$) be an orthogonal projector onto $\M^\bot$ (resp. $\Mp^\bot$). If $f$ contracts toward $\M$ with constant metric $\Theta$, then taking $U^T$ a partial isometry from $E_{\Mp^\bot}$ to $E_{\M^\bot}$, with kernel $0$ and image $Im(\Theta V V_2^T E_{\Mp})$, $f$ contracts toward $\Mp$ with constant metric $\Theta_2=U\Theta V V_2^T$
\end{lem}

See proof and definitions in appendix (\ref{proof:lemMall})
\begin{rem}
	\label{rem:lemMall_U}
	One would be tempted to set $U^T = \Theta V V_2^T$ but this is possible only when $\Theta$ itself is an orthonormal transformation. In particular when $\Theta=Id$ we can set $U^T=V V_2^T$ giving us $\Theta_2=Id$
\end{rem}

The main consequence of lemma \ref{lem:Mall} is the ability to
determine a sufficient contraction condition for a set $\Gamma$ of
symmetries. Indeed we just showed that a sufficient contraction
condition would be the contraction toward $\MG = \bigcap_{\g \in
\Gamma} \Mg $. This condition and the equivariance with respect to a
specific $\g$ allow lemma \ref{thm:flowcuong} to be applied to this $\g$.

In the generic linear case, there is no non trivial unifier, since the
intersection of all linear subspaces is reduced to $\{0\}$, corresponding to the
contraction of the full system.

But when considering the permutations, the common subspace exists :
$\Mall=\{\forall i, \forall j,\ x_i=x_j\}$. This subspace of full synchrony will be
very handy and may play an important role in neural networks.

Moreover the contraction toward $\Mall$ will be quite easy to prove when using
eventual symmetry of the system. To have more insight into the kind of
computation, see Section \ref{sec:permutdetails}.

In summary to the contraction and symmetry section, the main theorem we will use can be stated as
\begin{thm}
\label{thm:sym}
Consider a $\gamma$-equivariant or with $\Mg$ flow-invariant system, and assume
one
of the following contraction properties (sorted by decreasing strength)
	\begin{itemize}
		\item contraction of the system
		\item contraction toward $\MG$ with $\g\in\G$
		\item with $\gamma$ a permutation, contraction toward $\Mall$
	\end{itemize}
	Then all solutions converge exponentially to a  $\gamma$-symmetric
trajectory $x(t)$.
\end{thm}
\begin{proof}
	We only apply Lemma \ref{thm:flowcuong} and \ref{lem:cont_geq}.The only
change is the generalization in the fully contracting case by using $\Mg$
flow-invariance hypothesis : a contracting system has a unique solution
independent of the initial conditions, then taking a trajectory beginning in
$\Mg$ will stay in $\Mg$ by flow- invariance thus forcing the unique solution to
be in $\Mg$.
\end{proof}

\begin{rem}
All the symmetries of $f$, complying the theorem, will be transferred to the trajectories.\emph{The system will maximize synchrony }: for example, if $f$ complies the theorem with $\g=(0, 1)$ and $\g=(1, 2)$ then the system will lead to $x_0=x_1=x_2$.
\end{rem}

\section{Input continuity}
\label{sec:input_continuity}
\subsection{Input continuity motivations}

We now exploit these results on global symmetries in a control framework,
by introducing control inputs in the system to modulate its dynamics.

We will observe $f$ under the influence of some input $u$:
\begin{equation}
	\label{eq:generic_system}
	\dot{x}(t) = g(x,u(t),t) = f(x,t)
\end{equation}
The input $u$ doesn't represent all the input of the actual system, the rest of
the actual input can still be hidden as before inside the function $g$. This
choice underlines the fact that we want to study the system response to a
decisive part $u$ of the actual input. Then the idea is to control the output
with the input to get the property that we want see Section \ref{sec:control}.
In an in vivo situation, we can see all the feedback loops from the upper part
of the brain to the bottom part of the brain as control inputs $u$, but also the
input from the bottom to the top, and every other connections.

The response to $u$ of the system will be some output $s$ defined by the state
of the system. Typically we will use $s(t)=x(t)$ or some projection of the state
$s(t)=Px(t)$.

With this point of view we introduce {\it input continuity} analysis, in order
to describe the properties of a system's output knowing the properties of its
input while ensuring stability and robustness.

As a general matter, we want to be able to say two things, first ``if $u$ has
this property then $s$ will have that property'' and secondly ``if the input $u$
is close to an ideal input $u^\prime$ then $s$ will be close to the ideal
output $s^\prime$''. This will be formalized by the notion of
''input continuity'' :
\begin{align}
	\forall \epsilon \geq 0,\ \exists \eta \geq 0 \text{ such as }
d_1^t(u^\prime , u) \leq \eta \implies d_2^t(s^\prime , s) \leq \varepsilon
\label{eq:generic_input_continuity}
\end{align}
$d_1^t$ (resp. $d_2^t$) is a pseudo distance of the space of the input $u$
(resp. the output $s$). This pseudo distances help us to define the notion of
being close as traditionally but also the properties corresponding to the chosen
pseudo distance : if $d_1^t(u^\prime , u)=0$, then $u$ and $u^\prime$ are in the
same class defining  some property see Section \ref{sec:properties_distances}
for details with some interesting examples shown in Section
\ref{sec:codes_distances}.

\emph{Depending greatly on the distance we use, input continuity will be a
modular tool to ensure robustness of specific properties of the output given the
properties of the input.}

The modularity of input continuous block is something very generic and powerful,
more discussion can be found in section \ref{sec:lego_game}, but from now we
will restrict ourself to the study of a powerful input continuity found in
contracting systems.

\subsection{Input continuity and contraction}
\label{sec:contraction_inputcont}
There is no generic way to show input continuity of a
system. Depending on the type of system (discrete or continuous) and
the distance we use, we would have to examine each case we
encounter. But in the case of a contracting system, we can state some
powerful generic properties. This will allow us to combine Theorem \ref{thm:sym}
and the input continuity without any effort, the contraction being already an
hypothesis. The main tool to study input continuity of a contracting system is its robustness :

\subsubsection{Contracting systems}
We will compare the perturbed state $\sp$ with perturbed input $\up$
and the wanted state $s$ with perfect input $u$ :
\begin{align*}
	\dot{s} &= f(s,u,t)\\
	\dot{\sp} &= f(\sp,\up,t) = f(\sp,u,t) + h(\sp,t) \\
	\text{ with } h(t) &= f(\sp,\up,t) - f(\sp,u,t)
\end{align*}
Then, if $f$ is contracting, we can prove with the generalized form of the
robustness seen in ~\cite{Lohmiller1998} that
\begin{align}
	\label{robustness_contraction}
	\dot{R} + \lambda R \quad\leq&\quad\|h(t)\| \\
	\label{eq:m_R}
	R(t) \quad\leq &\quad e^{-\lambda (t-t_0)} R(t_0) +
\displaystyle\int_{t_0}^{t} e^{-\lambda (t-\tau)} \|h(\tau)\|\, d\tau \\
\nonumber
\end{align}
with $r = \sp - s$, $R(t) = \|r(t)\|$, $\|.\|$ the norm of the space in which
$f$ is proved contracting with contraction rate $\lambda$. By convention we will
have $R(-\infty) = 0$.
\begin{rem}
	If the contraction analysis uses a metric, it is reflected in the norm
$\|.\|$, for instance in the case of the use of the 2-norm $\|.\|_2$ and a
metric $\Theta$ we will use $\|.\|=\|\Theta\, .\|_2$
\end{rem}

We can prove the input continuity considering the space of the input signal and
output signal with these two norms :
\begin{align}
	\|x\|_{\infty}^{-\infty,t} &= Sup_{\ \tau < t} \{(\|x(\tau\|)\}
\label{norm_infiny} \\
	\Na(x) &= \int_{-\infty}^{t}\|x(\tau)\|e^{-\alpha (t-\tau)} \, d\tau \
\text{ with } \alpha>0 \label{eq:norm_alpha}
\end{align}

\begin{thm}
	If all the signals are bounded and $f$ is
contracting and uniformly continuous in time, then we have the input continuity
using the uniform norm (\ref{norm_infiny}) in both input and output space.
\end{thm}
\begin{proof}
	From \ref{robustness_contraction} we have : $\|r\|^{-\infty,t}_{\infty} \leq \frac{1}{\lambda}\|h\|^{-\infty,t}_{\infty}$. In the mean time, contraction gives us space continuity, then using the hypothesis of uniform continuity in time and the Heine's theorem over the compact space of bounded signals, $\exists k \in \mathbb{R},\ \|h\|^{-\infty,t}_{\infty} \leq k \|u^\prime -u\|$.
\end{proof}

Note that the boundedness of the input signals is a plausible condition {\it in vivo}.

In our control context, a more flexible and meaningful tool is the
norm (\ref{eq:norm_alpha}) with exponentially fast forgetting :
\begin{thm}
	\label{thm:continuity_contracting}
	If all the signals are bounded and $f$ is
contracting and uniformly continuous in time, then we have the input continuity
with $\Na$ in the input space and $\Nb$ in the output space.
\end{thm}
See proof in appendix \ref{proof:thm_continuity_contracting}. 

\begin{rem}
	When dealing with this kind of norms, Lemma \ref{lem:change_alpha} can be very convenient.
\end{rem}

\subsubsection{Systems contracting toward a subspace}
\label{sec:input_continuity_toward}
With contraction toward linear subspace, we want to ensure the property  ''we
are in $\M$`` which can be easily characterized with a generic semi-norm see
Section \ref{sec:properties_distances} for more explanations :
\begin{equation*}
	\label{eq:normM}
	N_{\mathcal{M}}(x) =  \|V x\|
\end{equation*}

We can now state an adapted version of Theorem \ref{thm:continuity_contracting}
\begin{thm}
	\label{thm:cont_cont_M}
	If all the signals are bounded, $f$ uniformly (in time) continuous, with Equation \ref{eq:cuong_sysy} contracting and $\mathcal{M}$ flow-invariant, then the system is input continuous with input norm $N_{\alpha,\|.\|}^{-\infty,t}$ and output pseudo norm $N_{\alpha,N_{\mathcal{M}}}^{-\infty,t}$.
\end{thm}
\begin{rem}
	This obviously works with the uniform norm in the same way.
\end{rem}
\begin{proof}
	We can apply Theorem (\ref{thm:continuity_contracting}) on the contracting system (\ref{eq:sys_M}) with space norm $\|.\|$, giving us input continuity of this system with norm $N_{\alpha,\|.\|}^{-\infty,t}$ ( resp. $N_{\beta,\|.\|}^{-\infty,t}$ ) as input norm ( resp. output norm ). Using notation of the Theorem, we can link $y$ and $x$, indeed if we set $y=Vx$ we get the system
	\begin{equation}
		\label{eq:sys_M}
		\dot{y}= V \dot{x} = Vf(x) = Vf(V^T y + U^T U x)
	\end{equation}
	which is contracting with respect to $y$, equivalently with the contraction of the system (\ref{eq:cuong_sysy}).Then we have $N_{\mathcal{M}}(x) = \|Vx\| = \|y\|$ so we can directly apply the result to the original system using $N_{\alpha,N_{\mathcal{M}}}^{-\infty,t}$ as output norm.
\end{proof}

\section{Control}
\label{sec:control}
With the power given by theorem \ref{thm:sym} and the flexibility given by input
continuity, we can now get the system to exhibit specific symmetries
with the help of a small controlling input. The global property of contraction
and input continuity will be required to robustly do transient and multiple
changes in the system and the symmetries of the output.

\subsection{Main idea}
Rather than looking at symmetrical solutions a system
may exhibit, as in the H/K Theorem of Golubitsky et
al.~\cite{Golubitsky1999,Buono2001,Golubitsky2000,Josic2006}, we
consider what symmetries the system may exhibit when submitted to
specific external inputs.

We consider a system of the form \[\dot{x}=f(x,t)=g(x,u(t),t)\]
where now $u(t)$ is a ``control input''. We will control the symmetry of the
system's output by modifying its input. To do so we will use the theorems
\ref{thm:sym} and \ref{thm:temporalsym} on the function $f$. We will need
\begin{itemize}
	\item a symmetry condition ($\gamma$-equivariance or flow-invariance)
	\item a contraction condition (contraction or contraction toward a
subspace)
\end{itemize}

The contraction condition (in any form) will give us input continuity
as shown in section \ref{sec:contraction_inputcont}, allowing us to
plug the input at any time instead of controlling the system from the
beginning, and still be exponentially close to the desired output. The
symmetry condition will lead the system to a state expressing the
desired symmetry.

The input can have different functions. It can determine the
contraction condition as explained in the following
\ref{sec:input_selection_of_contraction}, but also, and mainly, change
the symmetries of the system, as we now detail.

\subsection{Selection of spatio-temporal symmetries of the system}
\label{sec:selection_of_symmetry}
We first need to link the symmetries of $f$ and those of $g$ and
$u$. $g$ will be said h-equivariant if \[g(\gamma x,\gamma y,t) =
\gamma g(x,y,t+T)\]
\begin{thm}
	$g$ h-equivariant and $u$ h-symmetric $\implies$ $f$ h-equivariant
\end{thm}
\begin{proof}
	\begin{align*}
		f(\g x(t),t) &= g(\g x(t),u(t),t) = g(\g x(t),\g u(t+T),t) \quad
\text{ by symmetry of }u\\
		\g f(x(t),t+T) &= \g g(x(t),u(t+T),t+T) =g(\g x(t),\g u(t+T),t)
\quad \text{by equivariance of} g
	\end{align*}
\end{proof}

In the general case, $g$ h-equivariant is not sufficient, moreover,
increasing the symmetry thanks to the input is very unlikely, since it
would require an intelligent input, quite as complex as the neuron
model. This is not our goal since we consider the input to be
''small``, and the neuron model a realistic non linear dynamical
system. Thus for practical purposes the theorem is an
equivalence. Once the symmetries $\Gamma_g$ of $g$ are determined, we
can set an input with the symmetries $\Gamma_u$ to control the final
symmetries $\Gamma$ by considering $\Gamma = \Gamma_g \cap
\Gamma_u$, i.e., intersecting the symmetries of $g$ with those we set
in the input.

\begin{example} Consider the simple dynamics
\begin{equation*}
	f(x,t)=g(x,u(t),t) =
	\begin{cases}
		-x_1^3 + u_1(t) + sin(t)\\
		-x_2^3 + u_2(t) + sin(t+\pi/3)\\
		-x_3^3 + u_3(t) + sin(t+2\pi/3)
	\end{cases}
\end{equation*}
Here $g$ is $h=((1,2,3),\pi/3)$-equivariant, thus also
$h^2=((1,3,2),2\pi/3)$-equivariant and $h^3=(id,2\pi)$-equivariant, etc.
Taking for example $u$ $h^2$-symmetric, but not $h$-symmetric, we have $f$
$h^2$-equivariant but not $h$-equivariant. Note that then the solution will be
$6\pi$ periodic instead of $2\pi$ periodic, and also that nothing is needed or
proved about the respective phases of the different signals and elements.

This also shows that the creation of symmetries is unlikely. If
$g$ was only $h^2$-equivariant, having $u$ $h$-symmetric will not make
$f$ $h$-equivariant, but only $h^2$-equivariant.
\end{example}

\subsection{Control of the contraction condition of the system}
\label{sec:input_selection_of_contraction}
Having the system always contracting will probably not be the generic case and
the most biologically sound. Rather, we want to ``turn on'' the contraction
property at a specific time using the input. It can be represented by :

\begin{equation}
\label{eq:activator}
	\dot{x}=f(x,t) - k\chi_{on}x
\end{equation}
Having $k$ big enough and the activator $\chi_{on}=1$, the system will contract. With some systems like a set of FitzHugh-Nagumo elements, putting such a term only over the potential variables will give contraction toward $\Mall$.

This ``contracting input'' is a transient negative feedback loop which can
be turned on and off through the control of the activator. This can't be as
simple in neural models, but we propose a quite meaningful and simple
``implementation'' :

We consider the circuit of Figure \ref{fig:frequency_selector_system} but
without delay $d$. This circuit seems to be part of the neighborhood of many
cortical pyramidal cells in the treatment of extracortical afferent excitations
\cite{Buzaski2006}. We suggest a behavior : The pyramidal neuron $x_p$ would
have lots of gap junction with its touching inhibitory interneuron $x_i$. The
interneuron being way smaller will be driven by the pyramidal one, so that
$x_i\simeq x_p$. Next setting a low firing threshold for the interneuron would
allow to have it spiking proportionally to its potential. The resulting
inhibition of the pyramidal neuron will thus have the desired shape
$\simeq-k.x_p$. Input continuity permit to compute the distance between this
implementation and the perfect instantaneous negative feedback.

The activator will then be easily implemented by some inhibition of the
interneuron : no inhibition means $\chi_{on}=1$, $0$ otherwise.

The first two examples use the activated contraction to control spatio-temporal
symmetries (example of section \ref{ex:3ring} or section
\ref{ex:unstableinputcont}). Then we show in more details a grid example using
the $\Mall$ idea.

\subsection{Examples}
	Throughout this section we illustrate some of the above possibilities shown through basic examples, mostly using  FitzHugh-Nagumo neural models,
	\begin{align}
	\label{eq:FN}
		&\left\{
		\begin{aligned}
				\dot{v}&=v(\alpha -v)(v-1)-w+I\\
				\dot{w}&=\beta(v-\gamma w)
		\end{aligned}
		\right.\\
		&\alpha=6, \beta=3, \gamma=0.03 \nonumber
	\end{align}
with $I$ the synaptic input function.  Although most of the time we refer to
our system elements as ``neurons'', one should notice that more generally
the theory developed here applies to $f$ representing neural
networks, whose equations can be actually very
similar to FitzHugh-Nagumo models.

The use of FitzHugh-Nagumo neural models is motivated by its
simplicity while still a reasonably descriptive neuron model, and it has
the desired properties of contraction toward $\Mall$ when coupled only
through the potential variable see \cite{Pham2006}. This property is
kept when using the more precise Hodgkin-Huxley model see
\cite{Zyto2006}.

\subsubsection{Leading to unstable state : transient synchronization}
\label{ex:unstableinputcont}
\begin{figure}[!htb]
	\center
	\includegraphics[width=0.4\linewidth]{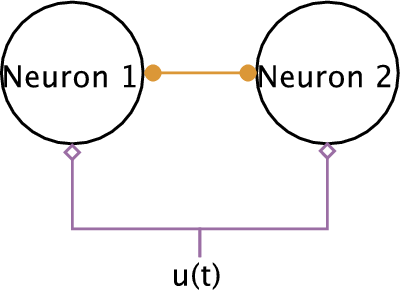}
	\caption{System for transient synchronization}
	\label{fig:syn_perfect_system}
\end{figure}

\begin{figure}[!htb]
	\center
	\subfloat[without noise]{\includegraphics[width=0.7\linewidth]{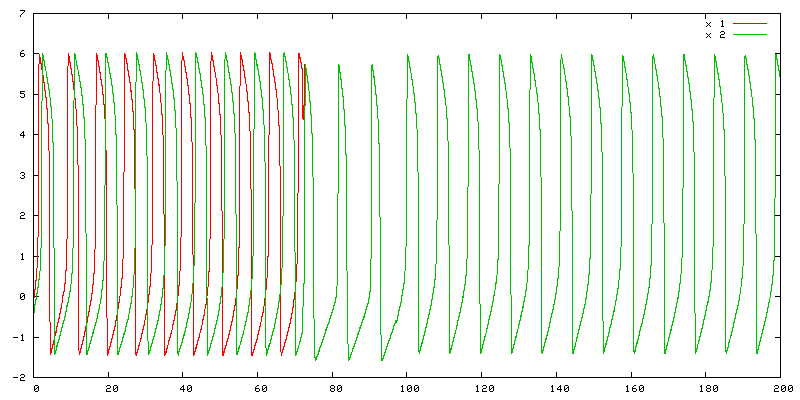}
\label{fig:syn_perfect_nonoise}}\\
	\subfloat[with noise]{\includegraphics[width=0.7\linewidth]{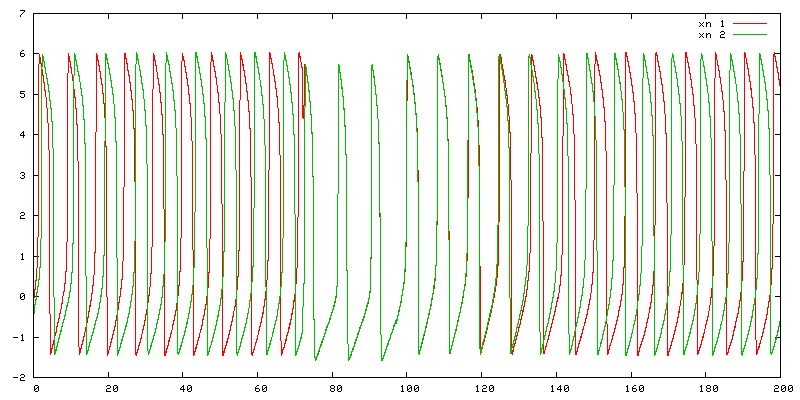}
\label{fig:syn_perfect_noise}}\\
	\subfloat[input $u$ with activator]{\includegraphics[width=0.7\linewidth]{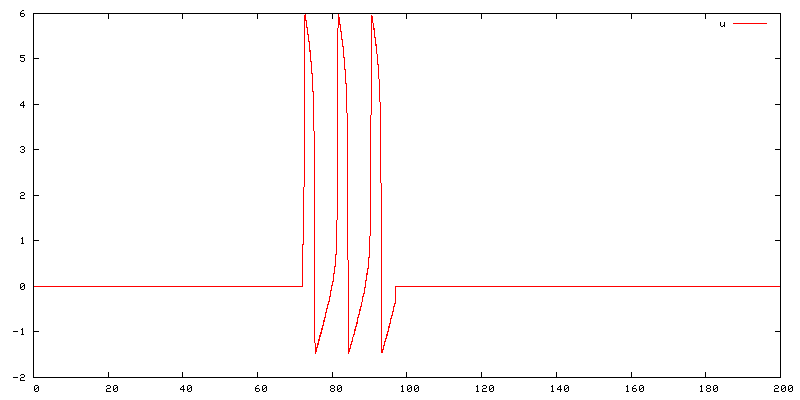}}
	\caption{Transient synchronization $\mu=0.1, \lambda=5, noise=5, e=20$ }
	\label{fig:syn_perfect_results}
\end{figure}

In this example we will use the action of a contracting input making
the system contracts toward $\Mall$ from time 75 to 95. This transient
contraction results in a transient synchronization, which is often
considered as a very important neural processing process
\cite{Palva2005,Palva2005a}. Consider the system seen in Figure
\ref{fig:syn_perfect_system}. Neuron 1 and 2 are two FitzHugh-Nagumo
neurons with an inhibitory symmetrical link between them. As we can
see Figure \ref{fig:syn_perfect_results} before we put the contracting
input, the mutual inhibitory link leads naturally the system to
antiphase. But after synchronizing the two neurons by force with the
input, they stay in the unstable state where they are equal see Figure
\ref{fig:syn_perfect_nonoise}, this during transient when some level
of noise is added see Figure \ref{fig:syn_perfect_noise}.

This example illustrates the idea that with the input we can lead the
system to a non 'natural' state, in much more complex networks this
could be some basic phenomenon to allow different computations with
the same network.

\begin{align*}
	\dot{x_1} &= f(x_1) -\mu x_2 + e + I_1 \\
	\dot{x_2} &= f(x_2) -\mu x_1 + e + I_2 \\
	I_i &= \lambda \chi_{t\in\lbrack75,95\rbrack}( u(t) - x_i)
\end{align*}

\clearpage
\subsubsection{Choose the spatio-temporal symmetry in a 3 ring hysteresis system}
\label{ex:3ring}
\begin{figure}[!htb]
	\includegraphics[width=0.4\linewidth]{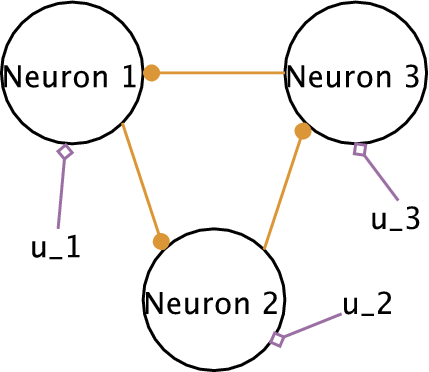}
	\caption{3 ring hysteresis system}
	\label{fig:3ring_system}
\end{figure}

\begin{figure}[!htb]
	\subfloat{\includegraphics[width=\linewidth]{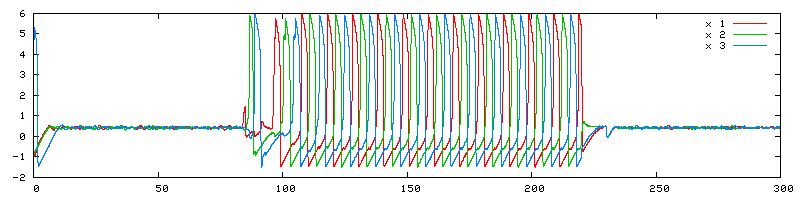}}
 \\
	\subfloat{\includegraphics[width=\linewidth]{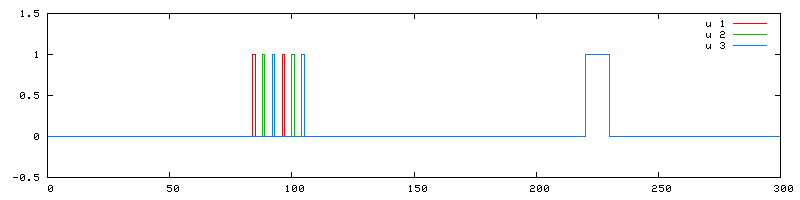}}

	\caption{3 ring hysteresis system, state selection by input}
	\label{fig:3ring_results}
\end{figure}

We have here Figure \ref{fig:3ring_system} a ring of 3 FitzHugh-Nagumo
neurons, each inhibiting its right neighbor. This system as we can
see Figure \ref{fig:3ring_results} has a stable state where none of the neurons
spike (here $e$ is not big enough to make them spike because of the
overall inhibition) but also another stable state $( (1,2,3) , T/3 )$
symmetric, where the neurons are spiking one after the other and each
neuron has a period $T$ (the inhibition being in the refractory period
of the next one, makes the FitzHugh-Nagumo spike shortly after). To pass
from one state to the other we use a contracting input function with
an input signal exhibiting the symmetry we want to see, in figure
\ref{fig:3ring_results} we lead the system to the rotating wave and
then back to silence.  The system
\begin{align*}
	\dot{x_1} &= f(x_1) -\mu x_3 + e_1 + I_1 \\
	\dot{x_2} &= f(x_2) -\mu x_1 + e_2 + I_2 \\
	\dot{x_3} &= f(x_2) -\mu x_2 + e_3 + I_2 \\
	I_i &= \lambda\ \chi_{on}( u_i(t) - x_i)
\end{align*}
\begin{rem}
	The control input to get back to the silent mode is here a long step but
can be any spatio-temporal identity signal ( equal for neuron 1, 2 and 3).
Inspired from the visual saccades involving bursting, this long step could model
a high frequency burst, being here a way to reinitialize our network to a silent
state before a new computation cf \cite{Kupper2005}.
\end{rem}

\clearpage
\subsubsection{Grid and group selection}

This example illustrate the idea of symmetry selection
without changing the contraction condition ( see section
\ref{sec:selection_of_symmetry}) and illustrating a common issue of 2D
segmentation. The concept will be to use a toric grid being
contracting toward $\Mall$, then to select thanks to
the input the desired (necessarily flow invariant) groups.

 The grid is formed with neurons connected through diffusive
connections ( representing gap junctions and other direct contacts
between neurons ) to their four closest neighbors. We will set the
coupling strength $k$ to be strong enough to ensure contraction toward
$\Mall$ ( it exists see balanced coupling in
\cite{Pham2006}).Following the control idea of section
\ref{sec:selection_of_symmetry}, the system will polysynchronize
depending on the flow invariant subspaces found in the input $u$.

\begin{figure}[!htb]
	\center
	\includegraphics[]{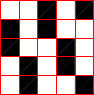}
	\qquad
	\includegraphics[]{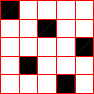}
	\caption{Grid pattern used to determine the groups}
	\label{fig:grid_patterns}
\end{figure}

\begin{figure}[!htb]
	\center
	\subfloat[Input]{\includegraphics[width=0.8\linewidth]{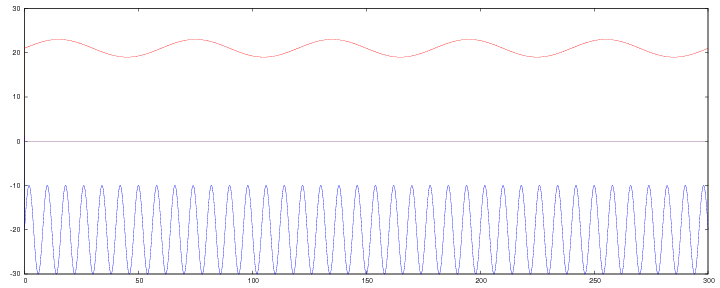}\label{fig:grid_inputs}}\\
	\subfloat[$k=0$]{\includegraphics[width=0.8\linewidth]{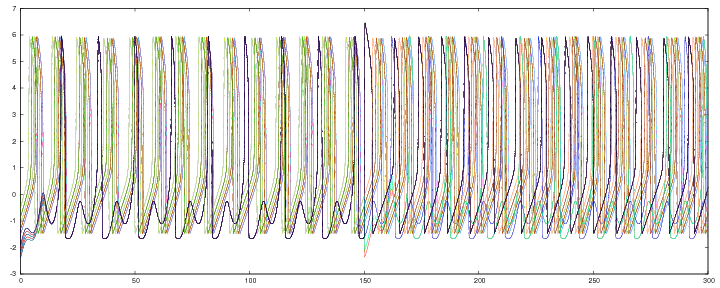}\label{fig:grid_0150}}\\
	\subfloat[$k=0.3$]{\includegraphics[width=0.8\linewidth]{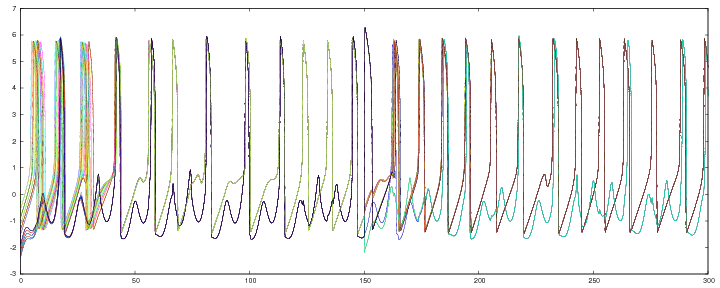}\label{fig:grid0.3150}}\\
	\subfloat[$k=0.7$]{\includegraphics[width=0.8\linewidth]{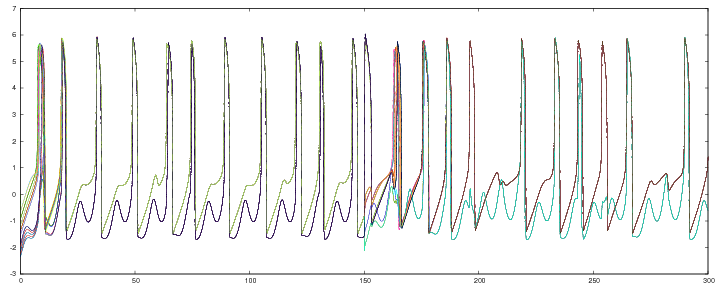}}
	\caption{Grid pattern 1 followed by pattern 2 at $t=150$}
	\label{fig:grid_s12_150}
\end{figure}

Specifically we use a 5x5 grid of identical FN neurons modeled as
(\ref{eq:FN}). For each run, the initial conditions are set so that
the neurons phase are spread out. We will consider 2 flow
invariant patterns chosen among the one we can find in
\cite{Antoneli2007}, namely pattern 1 and 2 of figure
\ref{fig:grid_patterns}, the coloring represents the wanted flow
invariant groups. $input\_0$ ( resp. $input\_1$ ) will represent the
input of the white (resp. black) neurons. To separate the two groups
of neurons but also showing some interesting interactions (we don't
want all the neurons to be synchronized even if the FitzHugh-Nagumo
model being a 2 dimensional model goes very easily to full synchrony
with this grid connection) we chose (see plots in figure
\ref{fig:grid_inputs})
\begin{align*}
	input\_0i = 2 \sin(2\pi t/60) +21 \\
	input\_1i = 10 \sin(2\pi t/8) - 20
\end{align*}
When we will say with noise we add to the input of each neuron a
random noise taking a new value between $0$ and $1$ each $0.05$
second.

We first apply pattern 1. Without coupling Figure \ref{fig:grid_0150} we observe
the natural behavior of the FitzHugh-Nagumo model which 'synchronize' with its
input quite easily as can be observed with one of the two groups. Setting
$k=0.3$ Figure \ref{fig:grid0.3150} gives the expected behavior, two groups
appear exponentially fast.

We then use the input to change the synchronized groups from pattern 1 to pattern
2 at time
$t=150$. First as expected the speed of convergence increase with $k$ but also
the
synchrony among groups, indeed the coupling tends to synchronize
groups also.

\begin{figure}[!htb]
	\center
	\subfloat[$k=0$]{\includegraphics[width=\linewidth]{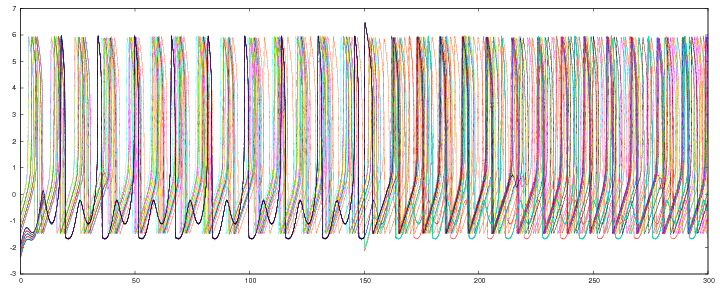}}\\
	\subfloat[$k=0.3$]{\includegraphics[width=\linewidth]{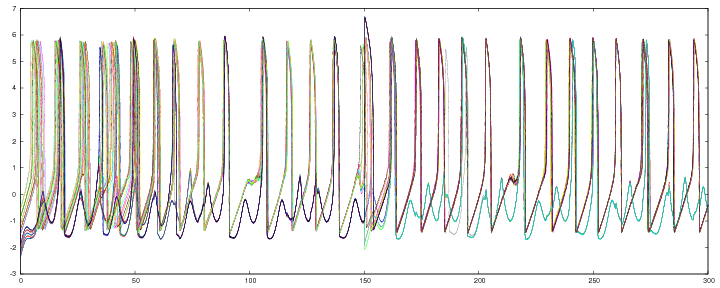}}\\
	\subfloat[$k=0.7$]{\includegraphics[width=\linewidth]{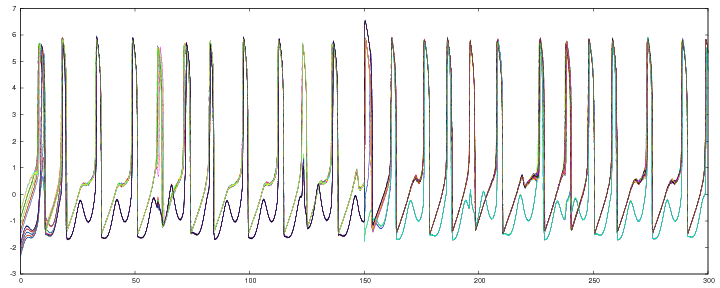}}
	\caption{Grid pattern 1 followed by pattern 2 at $t=150$ with noise}
	\label{fig:grid_s12_150_noise}
\end{figure}

Adding individual noise Figure \ref{fig:grid_s12_150_noise} obviously
prevent synchrony without the coupling, but with the coupling we
obtain the desired grouping with some glitches allowed by the input
continuity (the difference in the norms are always above a certain
mean of the integral of the noises).

\clearpage

\section{Discussion}

Robustness and globalness are interesting properties of the studied
methods. Many studies uses approximations about the trajectories,
considering that the neurons are close to their limit cycle, for
instance~\cite{Kopell2000,Kopell2004,Galan2005}. Since contraction is
a global property, nothing is assumed about the location of the state
of the system when we plug in a new input, allowing fast input-driven
switching between different synchronization patterns. Global
exponential convergence to the desired behaviors is obtained, with
quantifiable convergence rates.  Globalness also avoids some of the
topological difficulties associated to the study of large networks of
phase oscilllators. In the input continuity proof, only bounded
signals are needed.

The modularity of the tools is a strong property allowing to mix
studies of networks done at different scales (neuron, neural mass,
neural assemblies and so on).  Indeed, while we use neuron models as
our main dynamical system unit, the development can be applied to
other dynamical systems networks.

The symmetries used here are quite generic, the extension of spatial
symmetries to linear operators and the extension to spatio-temporal
symmetries seems important, since it is required to deal with the idea
of spatio-temporal pattern coding in the brain, and natural external
stimuli.

Two main weaknessness can be pointed out:

First our control over the symmetries of the system doesn't prevent
the system to exhibit {\it more} symmetries in the end $-$ mainly
ensuring to have two synchronized groups of neurons doesn't prevent to
have in fact total synchrony. To prevent the system to go to more
synchrony, it is important in practice to actively separate the groups
(as we did in the grid example), e.g. through inhibition, or break the
symmetry.

Second, the spatio-temporal case is very interesting and quite
unexplored. In this paper we only drew conclusions for fully
contracting systems, a restrictive condition. A relaxed condition
similar to the existence of $\Mall$ in the spatial case would be more
desriable (if perhaps unlikely).

Finally, the small circuit of a main neuron and its inhibitory
interneuron is interesting in its own right.  This circuit is proposed
as a plausible implementation of ``contracting inputs'' in section
\ref{sec:input_selection_of_contraction}, but also represents an
frequency selector circuit as seen in example
\ref{ex:frequency_selector}. Its biological relevance may be
further investigated.

\section{Appendix}
\subsection{Frequency selector / contraction activator network}
\label{ex:frequency_selector}
The small classical cortical circuit : Figure \ref{fig:frequency_selector_system} seems to be omnipresent on most cortical cells, set to treat the extracortical afferent excitations. \cite{Buzaski2006} We have already seen that this network could be of great use to control the contraction of the system Section \ref{sec:input_selection_of_contraction}. But with different parameters it can be an interesting frequency selector.

We will use the fact that we can control the frequency of both neurons with the input (by synchronizing the neuron with the input), conjugated with a fixed delay of inter-inhibition which will be the intrinsic frequency shut down of this circuit :

\begin{figure}[!htb]
	\centering
	\includegraphics[width=0.2\linewidth]{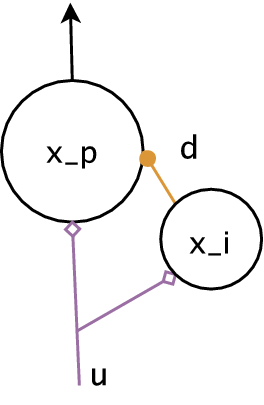}
	\caption{$d=11$, big enough $\lambda^{\prime}$}
	\label{fig:frequency_selector_system}
\end{figure}

\begin{figure}[!htb]
\begin{minipage}{0.5\linewidth}
	\subfloat{\includegraphics[width=0.8\linewidth]{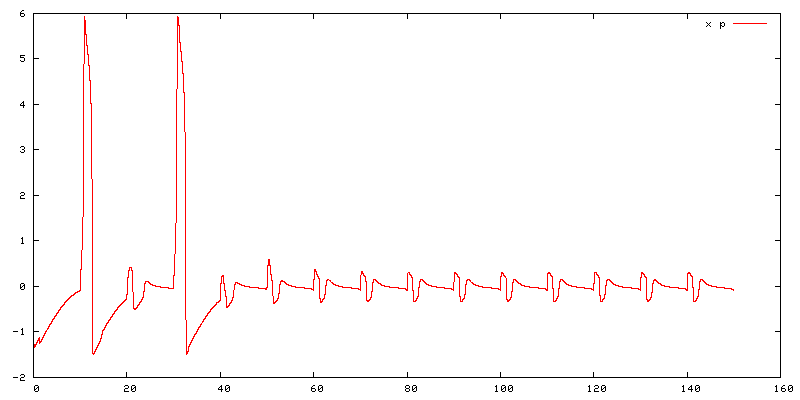}}\\
	\subfloat[$d=11$,$\lambda^{\prime}=1.3$, input period $10$]{\label{fig:freqSel_r1}\includegraphics[width=0.8\linewidth]{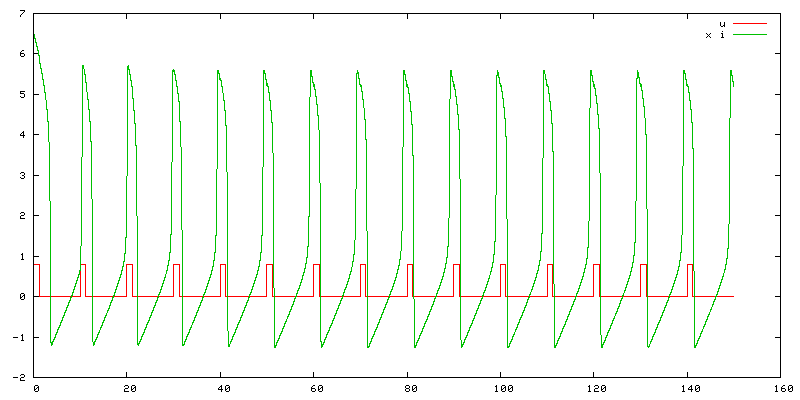}}
\end{minipage}
\begin{minipage}{0.5\linewidth}
	\subfloat{\includegraphics[width=0.8\linewidth]{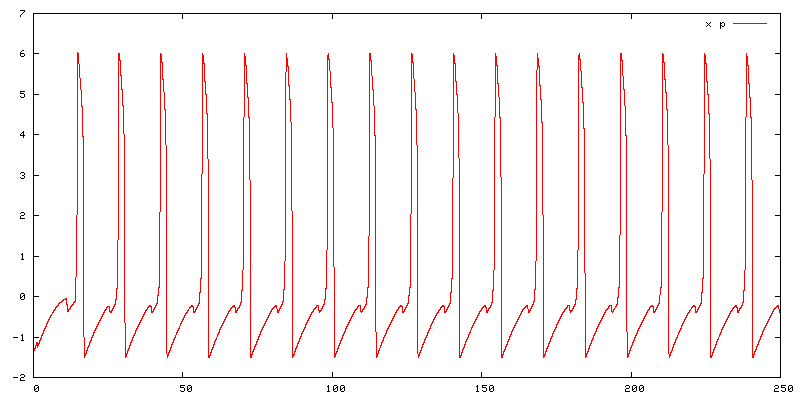}}\\
	\subfloat[$d=11$,$\lambda^{\prime}=1.3$, input period $14$]{\label{fig:freqSel_r2}\includegraphics[width=0.8\linewidth]{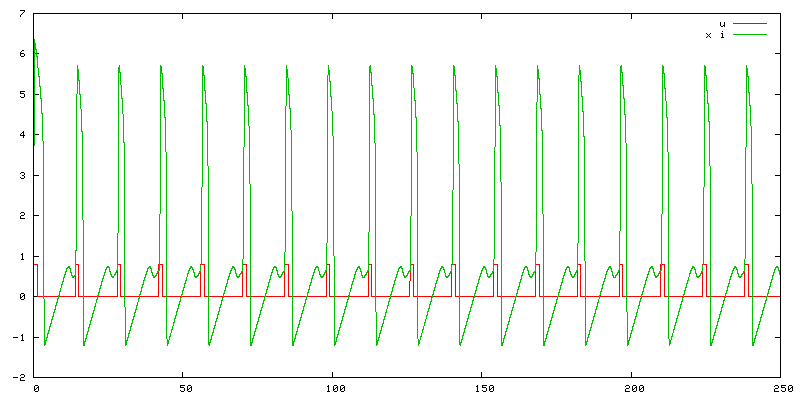}}
\end{minipage}
\end{figure}

There is a main pyramidal neuron $x_p$ connected to its inhibitory interneuron
$x_i$ with a synaptic delay $d$. We set an input $u$ with a specific frequency. First we use a frequency close to the corresponding delay, see Figure \ref{fig:freqSel_r1} we see $x_i$ which adapt to the input and then since the delay is of a close value, $x_p$ stops to spike. In Figure \ref{fig:freqSel_r2} we kept the delay of Figure \ref{fig:freqSel_r1} but set a further input frequency.

The system :
\begin{align*}
	\dot{x_p} &= f(x_p) + e + \lambda(u(t) - x_p) -w_i x_i(t-d) \\
	\dot{x_i} &= f(x_i) + e + \lambda^{\prime}(u(t) - x_i)
\end{align*}

\clearpage
\subsection{Input continuity precisions}

\subsubsection{Lego game}
\label{sec:lego_game}
The input continuity as introduced in Section \ref{sec:input_continuity} and
defined at (\ref{eq:generic_input_continuity}) allow us to play the Lego game :
plug in serial and parallel blocks having input continuity and get a bigger
block with input continuity. To plug in serial, the property of the output of
the first block should of course imply the property needed by the input of the
second block. The parallel block is just a redefinition of the input and input
space using for example as a new distance the $\sup$ of the two original
distances. Feedbacks are of a different kind of plug and we will need some more
refined analysis for example in two steps : block A has two input, one is a
feedback, if we can prove input continuity of A depending only on the first
input, we have some property of the output, then knowing that, we have some
property of the feedback input and we have a new ( and eventually stronger )
input continuity of A using the full property that we now know on the input,
giving the full. This kind of computation is close to the classical idea of
predictive top-down signal which is used to improve the treatment of the
feedforward input signal.

\subsubsection{Norms, distances and properties}
\label{sec:properties_distances}
There is a lot of different coding we can think of being used , like phase,
frequency, timing, spatial, etc ( some interesting examples among thousands
\cite{Leutgeb2005,Kupper2005} ) Each defining different ''distances`` between
signals and natural properties we could be tempted to prove on signals. In
general we will have a pseudo-distance in the signal space but a real distance
in the property space, with the equivalence classes $x \in [a] \iff d(x,a)=0$,
representing signals with the same property. To generalize this idea, we can
define a measure $\varphi^t :  \mathbb{R}^n \mapsto \mathbb{R}^m$ of our signal
which describe some properties that is if $\varphi^t(x_1) = \varphi^t(x_2)$ then
$x_1$ and $x_2$ have the same property at time t, with this we have a generic
pseudo-distance defined with the usual norm in the property space~:
\[d^t_{\varphi^t}(x_1,x_2) = \| \varphi^t(x_1) - \varphi^t(x_2) \|^t_m\]. This
construction was used for the input continuity of systems contracting toward a
linear subspace Section \ref{sec:input_continuity_toward}

\begin{rem}
	The interest of using a norm over using a distance : when is available a
norm coming from a dot product, we can define orthogonal projection on
subspaces, giving minimal distance between a real input and the space of desired
input.
\end{rem}
\subsubsection{Classical codes and distances}
\label{sec:codes_distances}
We can define several distances to represent classical coding or properties of
neuronal networks :
\begin{itemize}
	\item a pseudo-distance for frequency coding :\\
		Considering the signals defined by their spiking times :
$t^k_{x_1}$ and $t^k_{x_2}$, let $T^t_{x_1}$ be the mean distance between
$t^k_{x_1}$ and $t^{k+1}_{x_1}$ before time $t$ ( representing the mean period
of spiking ) then we could use :
		\[
		d^t(x_1,x_2)=|T^t_{x_1}-T^t_{x_2}|
		\]
	\item a pseudo-distance for synchrony in a certain subspace defined by a
projector $V$, remark that this is the one we mainly use throughout the paper:
		\[
		d^t(x_1,x_2)=|V(x_1)-V(x_2)|
		\]
\end{itemize}

Depending on the signal representation, we will have different classical
distances :
\begin{itemize}
	\item With signals as a set of spikes, described by the set of spike's
dates : to signal $u$  we associate $\{ \tau^u \}$ so that $u(t)= 1 \text{ if }
t \in \{ \tau^u \} \text{ else } 0$.
		We can then define :
		\[
		d^t(u,v) = \sum_{\tau_k \leq t \text{ in } \{\tau^u\} \cup
\{\tau^v\} } \left( |u(\tau_k) - v(\tau_k)|\right) \text{ with } \tau_k \in
\{\tau^u\}\cup \{\tau^v\}
		\]
		This one gives us a very precise distance between discrete
spatiotemporal patterns but is too sensitive.
		\begin{rem}
			Something important to notice is the fact that the
distances is defined at a time $t$, and may have access to the history of the
signal. The property can be varying with the time, since the distance is.
		\end{rem}

	\item We can define sensitivity delay $t_\varepsilon$ also seen as
refractory period, and use a kernel \cite{Gerstner2002} :
		\[ u(t) = \sum_{\tau^u_k \text{ in } \{\tau^u\}} K(t,\tau^u_k,
\{\tau^u\})\]
		\begin{itemize}
			\item  $K(.)= \delta(t - \tau^u_k)$ which gives a
formalization of the above vision,
			\item $K(.) = \delta(t - t_i^n)$ The spike train itself
$\sum_{t_i^n \in {\cal F}_i} \delta(t - t_i^n)$
			\item $K(.) = W_{t_i^{n-1}..t_i^n}(t) \,
\frac{t_\epsilon}{t_i^n - t_i^{n-1}}$  The normalized instantaneous frequency
({\scriptsize $t_i^n$ being predicted after $t_i^{n-1}$})
			\item $K(.) = \mbox{max}\left(0, \frac{t_\epsilon - |t -
t_i^n|}{t_\epsilon}\right)$  A non-causal measure of the instantaneous spike
density. \includegraphics[height=1cm]{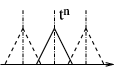}
			\item $K(.) = (1-e^\frac{-t_\epsilon}{\tau}) Y(t -
t_i^n) \, e^\frac{-(t - t_i^n)}{\tau}$  A $\tau$ time-constant, low-pass
filtered causal measure of.
			\item $K(.) = \mbox{ if } i = 0 \mbox { then }
\frac{t_\epsilon}{t_i^0 - t_\bullet} \mbox{ else } 0$  A representation of the
1st spike as in fast-brain mechanisms \cite{Thorpe2001}  with respect to a time
reference $t_\bullet$.
		\end{itemize}
		and then we can define instantaneous distances with
		\[ d^t(u,v) = | u(t)-v(t)|\]
		or define a more interesting distance as seen  in the following,
since thanks to the kernels we come back to real signal space.
	\item With the vision of spike set, we can consider the signals as
binary words (1 if the time is a spiking time, 0 otherwise), and use the usual
binary infinite word distances.
		This kind of norms can be useful if we are looking at some
binary coded properties, or to do some binary computation \cite{Carnell2005}.
Binary codes or barcodes \cite{Jin2004}.
	\item Some more statistical distances could be of interest, taking in
account the probability of spiking with respect to the history by some Hebbian
rule, for this we can be inspired by \cite{Aviel2003,Hayon2005,Toyoizumi2005}
	\item Phase synchrony measure giving also a pseudo distance inspired by
\cite{Pinsky1995}
\end{itemize}

Distances in real signal space : $\mathbf{R}^\mathbf{R}$, we can come back to
the usual functional norms, with $T$ the sensitivity window :
\[ \displaystyle N^t_2(u) = \int_{t-T}^{t} u(\tau)^2\, d\tau \qquad \text{ or }
N^t_1(u) = \int_{t-T}^{t} |u(\tau)|\, d\tau \]

or the convoluted of these ones, with $\mu \geq 0$
\begin{equation*} N^t_2(u) = \displaystyle\int_{t-T}^{t} u(\tau)^2 \mu(\tau-t)\,
d\tau \qquad \text{ or } N^t_1(u) = \displaystyle\int_{t-T}^{t} |u(\tau)|
\mu(\tau-t) \, d\tau \end{equation*}
And uses the norms as distances if we want with the usual : $d^t(u,v) =
N^t(u-v)$\\
We can list some properties of these norms :
\begin{itemize}
	\item growing with $T$ so the continuity with $T$ implies the one with
$T^{\prime} \geq T$
	\item $\mu$ will be very important to describe the system's sensitivity
	\item if we consider $u$ to be bounded it is sufficient to have $\mu
\sim\frac{1}{\tau^2} $ to allow to have an infinite window $T$
\end{itemize}
The norms are the generic case of the norm defined earlier Equation
\ref{eq:norm_alpha}.

\subsection{Contraction condition with help of symmetry}
\label{sec:permutdetails}
We have seen that with the contraction condition verified, we can
control the symmetry of the dynamics to control the
solution. Actually, the symmetry of the system can also help us to
simplify the proof of contraction itself.

\subsubsection{Circulant functions, symmetries and contraction conditions}

\begin{lem}
	A $\sigma$-equivariance with $\sigma$ a cycle of the size of the space
is something well known: $M$ $\sigma$-equivariant iff $M$ is circulant.\\
	\label{lem:circulant}
\end{lem}
\begin{proof}
	$M$ a $m \times m$ matrix is circulant iff $M$ seen as a quadratic form
$M(x,x) = x^T M x$ has the property $M(x, \sigma x) = M(\sigma^{-1}x,x)$ with
$\sigma$ a m-cycle.
	On the other side we have $\sigma^{-1} = \sigma^T$ and $M(x, \sigma x) =
 x^T M \sigma x = x^T \sigma M x = ( \sigma^{-1}x )^T M x$ by
$\sigma$-equivariance.
\end{proof}
\begin{rem}
	A matrix $C$ is circulant iff it can be written as :
	\[
	C =
	\begin{pmatrix}
		c(0) & c(1) & \dots & c(n-1) \\
		c(n-1) & c(0) & c(1) & \dots \\
		\ddots & \ddots & \ddots & \ddots \\
		\ddots & \ddots & \ddots & \ddots
	\end{pmatrix}
	\]
	The eigenvalue of the symmetric part are also simple :
	\[
	\mu_i = \sum_{k=0}^{n-1} c(k) cos(2ik\pi/n)
	\]
\end{rem}

We will consider a $\sigma$-equivariant (circulant cf \ref{lem:circulant})
system $\fp$ in space $E^\prime$ of size $n^\prime$.We define $\displaystyle
\lambda_j^\prime  = \sum_{k=0}^{n^\prime-1} \dfp{k}{0} cos(2jk\pi/n^\prime)$
\begin{rem}
	The $\lambda^\prime$ are defined using the first composant of $\fp$ but
thanks to the circulant property, it could be equivalently done using any other
composant
\end{rem}

Then we can prove two interesting lemmas :

\begin{lem}
	\label{lem:contcirc}
	\begin{equation}
		\label{eq:contcirc}
		\fp \text{ contracting with identity metric}  \iff \forall j \in
[0..\np-1] ,\,  \lambda_j^\prime < 0
	\end{equation}
\end{lem}
\begin{lem}
	\label{lem:vallfrom}
	\begin{equation}
		\label{eq:vallfrom}
		V_\sigma^{\prime} \frac{\partial \fp}{ \partial x}
V_\sigma^{\prime T} < 0 \iff \forall j \in [1..\np-1] ,\,  \lambda_j^\prime < 0
	\end{equation}
\end{lem}
\begin{rem}
	We have a condition of size $\np -1$ as it should be.
\end{rem}
\begin{rem}
	$V_{all}^\prime = V_\sigma^\prime$
\end{rem}
With this two lemmas we have moreover a simple link between the contraction and
the contraction toward the subspace of synchrony.

\subsubsection{Discrete symmetric system}

We consider $f$ a $\gamma$-equivariant system :
\begin{equation}
	\label{eq:hypo_gamma}
	f\gamma=\gamma f
\end{equation}

We will consider that $\gamma$ is a permutation and use notation from  Section
\ref{sec:discrete_operator}. $l_i$ is the size of $\sigma_i $, $E_{I_q}$ correspond to the space invariant by the action of $\g$ which is of dimension $q=n-\sum_i l_i$ since the cycles $\sigma_i$ are chosen non-trivial.

Such a $\gamma$-equivariant system will converge exponentially to a solution
$x=\gamma x$ if it has one of the contraction property. We will look at the
strongest one : contraction toward $\Mg =\{x=\gamma x\}$, and precisely at the
sufficient condition $\Vg \frac{\partial f}{ \partial x} \Vg^{T} < 0$ with $\Vg
x = 0 \iff x\in \Mg$.
\begin{thm}
	\[
	\Vg \frac{\partial f}{ \partial x} \Vg^{T} < 0 \iff \forall i\in [0..p]
\  \forall j \in \left[ 1 \dots l_i-1 \right], \ \lambda_{j,i} < 0
	\]
	with
	\begin{align*}
		\lambda_{j,i} &= \sum_{k=0}^{l_i-1} \partial_{k_i} f_{0_i}
cos(2jk\pi/l_i) &
	\end{align*}
	$k_i$ being the indice of the $k$th element in the $i$th subspace
(modulo $l_i$ to stay in the $i$th subspace). For example for $i=1$ we have
$k_1= l_1 + k$, $(-2)_1 = l_1 + l_2 -2$ and $f_{0_i}=f_{l_1}$
\end{thm}
\begin{proof}
\begin{rem}
	\label{rem:poly_multiuni}
	$x=\gamma x$ is equivalent to the conjunction $\forall i \in [0..p] \
x|_{E_{\sigma_i}}=\sigma_i \left( x|_{E_{\sigma_i}} \right) $. But
$x|_{E_{\sigma_i}}=\sigma_i \left( x|_{E_{\sigma_i}} \right) $ means that $x$ is
in synchrony inside each $E_{\sigma_i}$ : We want to prove a polysynchrony which
can be done by proving the synchrony inside each group.
\end{rem}
We can relate this remark to
\[
\Vg \frac{\partial f}{ \partial x} \Vg^{T} < 0 \iff \forall i \  V_{\sigma_i}
\left. \frac{\partial f}{ \partial x} \right| _{E_{\sigma_i}} {V_{\sigma_i}
}^{T} < 0
\]

Which leads us to us the result with Lemma \ref{lem:vallfrom}.
\end{proof}

\subsection{Proofs of Theorems, Lemma etc}

\subsubsection{Theorem \ref{thm:continuity_contracting}}
\begin{proof}
\label{proof:thm_continuity_contracting}
	We have the hypothesis:
	\begin{align*}
		\forall \tau \geq t_0,\ \|u(\tau)\| \leq M^\prime,\, \|s(\tau)\|
\leq M^\prime\\
		\NatO(\up - u) \leq \eta^\prime
	\end{align*}
	Using the uniform continuity,
	\begin{align}
		\forall \tau \geq t_0,\ \|h(\tau)\| \leq M \nonumber\\
		\Na(h) \leq \eta \label{hypothese1}
	\end{align}
	We have \[\Na(r) = \int_{-\infty}^{t}e^{\alpha(\tau-t)}\|r(\tau)\|\,d\tau\]
	However from the robustness (\ref{eq:m_R})\[ \forall \tau  \ \|r(\tau)\|
\leq \int_{-\infty}^{\tau} e^{\lambda(y-\tau)}\|h(y)\|\,dy\]
	We can apply Lemma \ref{lem:pre_boundary} with $t_0=-\infty$ and define
$t_1$ the moment of the saturation
	\begin{equation}
		e^{\lambda(t_1-t)} = \frac{\eta\lambda}{M}
		\label{eq:t_1_relation}
	\end{equation}
	\begin{align}
		\Na(r)
		&\leq \lim_{\,t_0 \to
-\infty}\int_{t_0}^{t_1}e^{\alpha(\tau-t)}\frac{M}{\alpha}\left( 1 -
e^{\lambda(t_0-\tau)} \right)\,d\tau + \int_{t_1}^{t} e^{\alpha(\tau-t)} \eta
e^{\lambda (t-\tau)}\,d\tau \notag \\
		\intertext{which by calculus using the relation Equation
\ref{eq:t_1_relation}}
		&\leq \frac{M}{\lambda\alpha}\left( \frac{\eta\lambda}{M}
\right)^{\frac{\alpha}{\lambda}} + \frac{\eta}{\alpha-\lambda}\left(
1-\left(\frac{\eta\lambda}{M} \right)^{\frac{\alpha-\lambda}{\lambda}}\right)
		\label{eq:majoration_continuite_infiny}
	\end{align}

	From (\ref{eq:majoration_continuite_infiny}) we have the continuity,
since all the powers of $\eta$ are positives.
\end{proof}

Let's consider some cases :
\begin{itemize}
	\item if $\alpha \leq \lambda$ We first have to say that we still have a
positive term, since the exponent change also its sign, making $\left(
1-\left(\frac{\eta\lambda}{M} \right)^{\frac{\alpha-\lambda}{\lambda}}\right)
\leq 0$ then the fact that $M$ is finite is important, otherwise this term will
go to $+ \infty$ and we can have in extreme cases some numerical surprises (even
if we have continuity).
	\item if $\alpha = \lambda$ then by continuity we get the limit : $
\frac{\eta}{\alpha} + \frac{\eta}{\alpha}ln\left( \frac{M}{\eta\alpha}\right) $
so $M$ should still be bounded to allow us to use this limit..
	\item if $\alpha \geq \lambda$ then we can get rid of $M$ and perhaps in
first approximation, just keep the main term : $\frac{\eta}{\alpha-\lambda}$
\end{itemize}

\begin{rem}
	If $t_1$ doesn't exist ( i.e. $ \frac{\eta\lambda}{M} \geq 1$ ) it first
means that we did not took $\eta$ small but the calculus gives just the first
term of (\ref{eq:majoration_continuite_infiny}) which is still good.
\end{rem}
\begin{rem}
	We can also instead of using some case based calculus over the $\lambda$
and $\alpha$ use the lemma \ref{lem:change_alpha} to have a pseudo equivalence
of all of these norms.
\end{rem}
\begin{rem}
	We should see that in the preceding proof, $\lambda$ is taken as the
contraction rate of the system, but we can use any $\acute{\lambda} \leq
\lambda$ since (\ref{robustness_contraction}) will still be true.
\end{rem}

\begin{lem}
	\label{lem:pre_boundary}
	a boundary on the norm before $t$:
	\begin{equation}
		\label{eq:first_property}
		\forall t \geq \tau \geq t_0,\
\int_{t_0}^{\tau}e^{\beta(y-\tau)}H(y)\,dy \leq min\left( \eta e^{\beta
(t-\tau)}\ ,\ \frac{M}{\beta}(1 - e^{\beta (t_0-\tau)}) \right)
	\end{equation}
	Remark : The inequality is an equality for
	\[
	H(y) =
	\begin{cases}
		M & \text{ if $y \leq t_1$,}\\
		0 & \text{ if $t_1 \leq y \leq t$ }
	\end{cases}
	\]
	with $t_1$ the moment of saturation if it exists :
	\[
	\frac{M}{\beta}(e^{\beta(t_1 - t)} - e^{\beta(t_0 - t)}) = \eta
	\]
\end{lem}
\begin{proof}
	\begin{align*}
		\forall \tau \geq t_0,
\int_{t_0}^{\tau}e^{\alpha(y-\tau)}H(y)\,dy
		&= e^{\alpha(t-\tau)} \int_{t_0}^{\tau}e^{\alpha(y-t)}H(y)\,dy\\
		&\stackrel{\text{\tiny def}}{=} e^{\alpha(t-\tau)} h(\tau)
	\end{align*}
	however
	$h(t_0)=0,\quad h(t)=\eta,\quad \dot{h}(\tau)=e^{\alpha(\tau-t)}H(y)$
	\begin{alignat*}{5}
		\text{from (\ref{hypothese1}) } \quad& 0 &\quad& \leq &\quad&
\dot{h}(\tau) &\quad& \leq &\quad& M e^{\alpha(\tau-t)} \\
		& 0 && \leq && h(\tau) && \leq && \min\left( \eta \ ,\
\int_{t_0}^{\tau}e^{\alpha(y-t)}M\,dy \right) \\
		&0 && \leq && h(\tau) && \leq && \min\left( \eta \ ,\ \frac{M
e^{-\alpha t}}{\alpha}(e^{\alpha\tau} - e^{\alpha t_0}) \right)
	\end{alignat*}
\end{proof}

\begin{lem}
	\label{lem:change_alpha}
	The possibility to change the $\alpha$ of the norm keeping a boundary:
	\begin{eqnarray}
		\label{eq:second_property}
		\Nb(d) \leq q(\eta)\\
		\text{ with } &q(\eta) \leq \eta &\text{ if } \alpha \leq \beta
\nonumber \\
		\text{ and } &q(\eta) \leq \frac{M}{\beta}\left( \frac{\eta
\alpha}{M} \right)^{\frac{\beta}{\alpha}} &\text{ if } \beta \leq \alpha
\nonumber
	\end{eqnarray}
	Remark : it is also true that we keep the boundary with $t_0 \neq
-\infty$ but the result is less interesting.
\end{lem}
\begin{proof}
	With the notation of lemma \ref{lem:pre_boundary} :
	\begin{align*}
		\Nb(d)
		&= \int_{t_0}^{t}e^{\beta(\tau-t)}D(\tau)\,d\tau \\
		&= e^{(\alpha-\beta)t}
\int_{t_0}^{t}\dot{h}(\tau)e^{(\beta-\alpha)\tau}\,d\tau \\*
		\intertext{integrating by parts and taking the limit
$t_0=-\infty$ when possible :}
		&= h(t) + (\alpha -\beta)
\int_{t_0}^{t}h(\tau)e^{(\beta-\alpha)(\tau-t)}\,d\tau\\*
		\intertext{for $\alpha \geq \lambda$ and using the property of
$h(\tau)$ seen in lemma \ref{lem:pre_boundary} }
		&\leq \eta + (\alpha - \beta) \left[ \int_{t_0}^{t_1}\frac{M
e^{-\alpha t}}{\alpha}(e^{\alpha\tau} - e^{\alpha
t_0})e^{(\beta-\alpha)(\tau-t)} + \int_{t_1}^{t}\eta
e^{(\beta-\alpha)(\tau-t)}\right]\\
		\intertext{which gives us, with $t_0=-\infty$ ( which works
without any new hypothesis)}
		&\leq  \eta + (\alpha - \beta) \left[ \frac{M
e^{\beta(t_1-t)}}{\alpha\beta} + \frac{\eta}{\beta - \alpha}(1 -
e^{(\beta-\alpha)(t_1-t)}) \right]\\
		\intertext{ and using the relation \eqref{eq:t_1_relation} }
		&\leq \frac{M}{\beta}\left( \frac{\eta\alpha}{M}
\right)^{\frac{\beta}{\alpha}}
	\end{align*}
\end{proof}

\subsection{Proof of Lemma \ref{lem:vallfrom}}

\begin{proof}

	A natural projector to the subspace of synchrony is $W$:
	\begin{align*}
		\begin{aligned}
			W =  I - \sigma =
			\begin{pmatrix}
				1 & -1 & 0 & \dots & 0 \\
				0 & \ddots & \ddots & \ddots & \vdots \\
				\vdots & \ddots & \ddots & \ddots & 0 \\
				0 & \ddots & \ddots & \ddots & -1\\
				-1 & 0 & \dots & 0 & 1
			\end{pmatrix}\\
		\end{aligned}
	\end{align*}

	To use our natural projector, we first have to remark that it
	is not really a projector to the orthogonal space. Such a projector
	can be obtained by removing the redundancy inside $W$, because
	it is a system of dimension $n$ but represent an hyperspace
	of dimension $n-1$.

	Then this noticed, instead of looking for the condition $\Vp
\frac{\partial f}{ \partial x} \Vpt < 0$ we will look for the equivalent
condition $W \frac{\partial f}{ \partial x} W^T $ strictly negative except for
one null eigenvalue.

	\begin{align*}
		A &= W \frac{\partial \fp}{ \partial x} W  \\
		&= \left( \dfp{s}{r} - \dfp{s+1}{r} - (\dfp{s}{r+1} -
\dfp{s+1}{r+1})  \right)_{(r,s)\in [0..\np-1]^2} \\
		&= \left( 2 \dfp{s}{r} - \dfp{s-1}{r} - \dfp{s+1}{r}
\right)_{(r,s)\in [0..\np-1]^2} \text{ using that $\partial \fp$ is circulant}
	\end{align*}

	\begin{rem}
		All indices are modulo $\np$.
	\end{rem}
	\begin{rem}
		$W$ is circulant.
	\end{rem}

	Since $A$ is circulant (product of circulant matrix), it is defined by
	\[
	a(k) = A_{0,k} = 2 \dfp{k}{0} - \dfp{k-1}{0} - \dfp{k+1}{0}
	\]
	We are interested in $C$ its symmetric part. It is also a circulant
matrix, so defined by $c(k) \ k\in[0..l_i-1]$ :
	\[
	c(k) = \frac{a(k)+a(-k)}{2} = \frac{1}{2}\left(2 \dfp{k}{0}  -
\dfp{k-1}{0} - \dfp{k+1}{0} + 2 \dfp{-k}{0} - \dfp{-(k-1)}{0} - \dfp{-(k+1)}{0}
\right)
	\]
	One of the interest of circulant matrix is that we know their eigenvectors
	and associated eigenvalues : taking
	$\wp_j = e^{\mathbf{i} \frac{j 2 \pi}{\np}}$ one of the $\np$th root of 1
	( ${\wp_j}^{\np}=1$), we construct the eigenvector
	$\displaystyle v_j=( 1, \wp_j ,{\wp_j}^2, \dots, {\wp_j}^{\np-1})$
	associated to the eigenvalue
	$\displaystyle \mu_j^\prime= \sum_{k=0}^{\np-1} c(k){\wp_j}^k$\\
	\begin{rem}
		Since we took $C$ symmetric ($c(k) = c(-k )$), the eigenvalues
$\mu_j^\prime$ are all real as one could remark regrouping $ c(k){\wp_j}^k +
c(-k){\wp_j}^{-k} = c(k)cos(2jk\pi/\np)$.
	\end{rem}

	With some simple regrouping  :

	\begin{align*}
		\mu_j^\prime &= \sum_{k=0}^{\np-1} c(k){\wp_j}^k = 2 (1 -
({\wp_j} + {\wp_j}^{-1})/2 ) \lambda_j^\prime = 2 (1 - cos(2j\pi/l_0))
\lambda_j^\prime \\
		\text{with } \lambda_j^\prime &= \sum_{k=0}^{\np-1} \partial_{k}
f_0 ({\wp_j}^k + {\wp_j}^{-k})/2 = \sum_{k=0}^{\np-1} \partial_{k} f_0
cos(2jk\pi/\np)
	\end{align*}

	And :
	\begin{align*}
		&\mu_0^\prime = 0\\
		\forall j \in [1..\np-1],\qquad & sign(\lambda_j^\prime) = sign
(\mu_j^\prime)
	\end{align*}

	Since $W$ is of dimension $\np$, but represent a $\np-1$ dimensional
space, corresponding to $\mu_0^\prime$. Then by virtue of Theorem 1 of
\cite{Pham2006} the contraction to the subspace of synchrony $E^\prime_{all}$ is
ensured with $\forall j \in [1..\np-1] ,\, \mu_j^\prime < 0$ or equivalently
$\forall j \in [1..\np-1] ,\, \lambda_j^\prime < 0$
\end{proof}

\subsection{Contraction toward $\Mall$ using symmetry}
\subsubsection{Proof of Lemma \ref{lem:Mall}}
Let's recall the definition of a partial isometry : it is an isometry from the orthogonal of its kernel to its image.
\begin{proof}
\label{proof:lemMall}
We take $VV^T = I_{E_{\M^\bot}}$, $V_2V_2^T = I_{E_{\Mp^\bot}}$, and $\forall x \in E_{\M^\bot},\ x^T\Theta V f V^T \Theta^{-1}x<0$.

Note first that $V_2V^T\Theta^T\Theta V V_2^T > {\bf 0}$. Indeed, $\Theta$ is invertible and $\forall y \in E_{\Mp^\bot},\ VV_2^Ty = 0$ implies $\ V_2^Ty = 0$ (since $\Mp^\bot \subset \M^\bot$), which in turn implies $\ y = 0$ (by definition).

Next note that \[ \exists \Theta_2 \in GL(E_{\Mp^\bot}),\ \forall y \in E_{\Mp},\ \exists x \in E_{\M},\ V^T \Theta^{-1}x = V_2^T \Theta_2^{-1} y \text{ and } V^T \Theta^Tx = V_2^T \Theta_2^T y\] Indeed, since $VV^T=I_{E_{\M^\bot}}$ this is equivalent to :
	\begin{align*}
		\exists \Theta_2 \in GL_m&,\ \forall y \in E_{\Mp^\bot},\ &\Theta V V_2^T\Theta_2^{-1} y &= (\Theta^T)^{-1} V V_2^T \Theta_2^Ty \\
		\exists \Theta_2 \in GL_m&,\ &V_2V^T\Theta^T\Theta V V_2^T  &= \Theta_2^T\Theta_2
	\end{align*}
where $m$ is the dimension if the subspace $E_{\Mp^\bot}$, $\Theta_2 $ exists because $V_2 V^T \Theta^T \Theta V V_2^T > {\bf 0}$.

Finally, by unitary freedom of square roots for symmetric positive operators there exists a partial isometry $U$ such that $\Theta_2 = U \Theta V V_2^T$. Thus, \[\Theta_2 V_2 f V_2^T \Theta_2^{-1} = U \Theta V V_2^T V_2 f V_2^T (U \Theta V V_2^T)^{-1} = U \Theta V f V^T \Theta^{-1} U^T<0 \] where the last expression is negative definite from the hypothesis.

As is $\Theta_2$, $U$ is defined up to isometries. We can thus take any $U$ having the sufficient following properties. $U^T$ is basically the partial isometry embedding $E_{\Mp}$ in $Im(\Theta V V_2^T E_{\Mp})$ which is of the same dimension but in a bigger space : $U^TU$ projects $E$ onto $Im(\Theta V V_2^T E_{\Mp})$ and $UU^T=I_{E_{\Mp}}$. \end{proof}

\subsubsection{Proof of Lemma \ref{lem:Mall2}}
\begin{proof}
\label{proof:lemMall2}
	Using a similar construction as in proof \ref{proof:lemMall}, with $m$
the dimension of the subspace $E_{\M^\bot}$, we need
	\begin{align*}
		\exists \Theta_2 \in GL_m,\quad \forall y \in E_{\M^\bot},
\Theta V V_2^T\Theta_2^{-1} y &= (\Theta^T)^{-1} V V_2^T \Theta_2^Ty
	\end{align*}
	which is equivalent to :
	\begin{align*}
		\exists \Theta_2 \in GL_m,\quad (V_2^T)^{-1}V^T\Theta^T\Theta V
V_2^T  &= \Theta_2^T\Theta_2 \\
		(T^T)^{-1}\Theta^T\Theta T^T  &= \Theta_2^T\Theta_2
	\end{align*}
	things works with $T$ unitary, but if not unitary there are few chances
for i to work.
\end{proof}

\bibliographystyle{unsrt}

\bibliography{biblio}

\begin{thebibliography}{10}

\bibitem{Mountcastle1978}
Vernon~B. Mountcastle.
\newblock {\em An Organizing Principle for Cerebral Function: The Unit Model
  and the Distributed System and The Mindful Brain}.
\newblock Gerald M. Edelman and Vernon B. Mountcastle, 1978.

\bibitem{Koerner1999}
E.~Körner, M.-O. Gewaltig, U.~Körner, A.~Richter, and T.~Rodemann.
\newblock A model of computation in neocortical architecture.
\newblock {\em Neural Networks}, 12:989--1005, 1999.

\bibitem{Kupper2006}
Rüdger Kupper, Andreas Knoblauch, Marc-Oliver Gewaltig, Ursula Körner, and
  Edgar Körner.
\newblock Simulations of columnar architecture for cortical stimulus
  processing.
\newblock 2006.

\bibitem{Milo2002}
R.~Milo, S.~Shen-Orr, S.~Itzkovitz, N.~Kashtan, D.~Chklovskii, and U.~Alon.
\newblock Network motif: Simple building blocks of complex networks.
\newblock {\em SCIENCE}, 298:824--827, october 2002.

\bibitem{Hayon2005}
Gaby Hayon, Moshe Abeles, and Daniel Lehmann.
\newblock A model for representing the dynamics of a system of synfire chains.
\newblock {\em Journal of Computational Neuroscience}, 18:41--53, 2005.

\bibitem{Malsburg1995}
C.~von~der Malsburg.
\newblock Binding in models of perception and brain function.
\newblock {\em Curr Opin Neurobiol}, 5(4):520–526, Aug 1995.

\bibitem{Robert1993}
GÃŒtig Robert.
\newblock Learning input correlations through nonlinear temporally asymmetric
  hebbian plasticity.
\newblock {\em Journal of Neuroscience}, 23(9):3697--3714, 1993.

\bibitem{Bienenstock1982}
E.~L. Bienenstock, L.~N. Cooper, and P.~W. Munro.
\newblock Theory for the development of neuron selectivity: orientation
  specificity and binocular interaction in visual cortex.
\newblock {\em J Neurosci}, 2(1):32--48, Jan 1982.

\bibitem{Abarbanel2002}
Henry D~I Abarbanel, R.~Huerta, and M.~I. Rabinovich.
\newblock Dynamical model of long-term synaptic plasticity.
\newblock {\em Proc Natl Acad Sci U S A}, 99(15):10132--10137, Jul 2002.

\bibitem{Malenka2004}
Malenka and Bear.
\newblock Ltp and ltd: an embarrassment of riches.
\newblock {\em Neuron}, 44(1):5--21, sept 2004.

\bibitem{Izhikevich2004}
E.M. Izhikevich.
\newblock Which model to use for cortical spiking neurons?
\newblock {\em Neural Networks, IEEE Transactions on}, 15(5):1063–1070,
  September 2004.

\bibitem{Sherman2002}
S.~Murray Sherman and R.~W. Guillery.
\newblock The role of the thalamus in the flow of information to the cortex.
\newblock {\em Philos Trans R Soc Lond B Biol Sci}, 357(1428):1695–1708, Dec
  2002.

\bibitem{Binzegger2004}
Tom Binzegger, Rodney~J Douglas, and Kevan A~C Martin.
\newblock A quantitative map of the circuit of cat primary visual cortex.
\newblock {\em J Neurosci}, 24(39):8441–8453, Sep 2004.

\bibitem{Kennedy2007}
Colette Dehay and Henry Kennedy.
\newblock Cell-cycle control and cortical development.
\newblock {\em Nat Rev Neurosci}, 8(6):438–50, June 2007.

\bibitem{Lohmiller1998}
Winfried Lohmiller and Jean-Jacques Slotine.
\newblock On contraction analysis for nonlinear systems.
\newblock {\em Automatica}, 34(6):617--682, 1998.

\bibitem{Spong2005}
M.W. Spong and F.~Bullo.
\newblock Controlled symmetries and passive walking.
\newblock {\em Automatic Control, IEEE Transactions on}, 50(7):1025–1031,
  July 2005.

\bibitem{Golubitsky1999}
M.~Golubitsky, I.~Stewart, P.L. Buono, and J.J. Collins.
\newblock Symmetry in locomotor central pattern generators and animal gaits.
\newblock {\em Nature}, 401:693--695, 1999.

\bibitem{Boergers2003}
Christoph Börgers and Nancy Kopell.
\newblock Synchronization in network of excitatory and inhibitory neurons with
  sparse, random connectivity.
\newblock {\em Neural Computation}, 15:509--538, 2003.

\bibitem{Pinsky1995}
Paul~F. Pinsky and John Rinzel.
\newblock Synchrony measures for biological neural networks.
\newblock {\em Biological Cybernetics}, 73:129--137, 1995.

\bibitem{Kazanovich2003}
Yakov Kazanovich and Roman Borisyuk.
\newblock Synchronization in oscillator systems with a central element and
  phase shifts.
\newblock {\em Progress of Theoretical Physics}, 110(6):1047--1057, december
  2003.

\bibitem{Lohmiller2000}
Winfried Lohmiller and Jean-Jaques~E. Slotine.
\newblock Nonlinear process control using contraction theory.
\newblock {\em A.I.Ch.E. Journal}, 46(3):588--597, 2000.

\bibitem{Slotine2002}
Jean-Jacques~E. Slotine.
\newblock Modular stability tools for distributed computation and control.
\newblock {\em Int J. Adaptative Control and Signal Processing}, 17(6), october
  2002.

\bibitem{Wang2004}
Wei Wang and Jean-Jacques~E. Slotine.
\newblock On partial contraction analysis for coupled nonlinear oscillator.
\newblock {\em Biological Cybernetic}, 92(1), 2004.

\bibitem{Pham2006}
Quang-Cuong Pham and Jean-Jacques~E. Slotine.
\newblock Stable concurrent synchronization in dynamic system networks.
\newblock {\em Neural Netw}, Oct 2006.

\bibitem{Buono2001}
Pietro-Luciano Buono and Martin Golubitsky.
\newblock Models of central pattern generators for quadruped locomotion.
\newblock {\em J. Math. Biol.}, 42:291--326, 2001.

\bibitem{Golubitsky2000}
Martin Golubitsky and Ian Stewart.
\newblock Patterns of oscillation in coupled cell systems.
\newblock August 2000.

\bibitem{Josic2006}
Kresimir Josic and Andrei Török.
\newblock Network architecture and spatio-temporally symmetric dynamics.
\newblock may 2006.

\bibitem{Buzaski2006}
Buzaski.
\newblock {\em Rhythms of the brain}.
\newblock New York : Oxford University Press, 2006.

\bibitem{Zyto2006}
Sacha Zyto and Jean-Jaques~E. Slotine.
\newblock Contraction analysis of conductance-based oscillators.
\newblock Technical report, MIT Nonlinear Systems Laboratory, May 2006.

\bibitem{Palva2005}
J.~Matias Palva, Satu Palva, and Kai Kaila.
\newblock Phase synchrony among neuronal oscillations in the human cortex.
\newblock {\em J Neurosci}, 25(15):3962–3972, Apr 2005.

\bibitem{Palva2005a}
Satu Palva, Klaus Linkenkaer-Hansen, Risto Naatanen, and J.~Matias Palva.
\newblock Early neural correlates of conscious somatosensory perception.
\newblock {\em J. Neurosci.}, 25(21):5248--5258, 2005.

\bibitem{Kupper2005}
Rüdiger Kupper, Marc-Olivier Gewaltig, Ursula Körner, and Edgar Körner.
\newblock Spike-latency codes and the effects of saccades.
\newblock {\em Neurocomputing}, 65(66):189--194, december 2005.

\bibitem{Antoneli2007}
F.~Antoneli, A.P.S. Dias, M.~Golubitsky, and Y.~Wang.
\newblock Synchrony in lattice differential equations.
\newblock In {\em Some Topics In Industrial and Applied Mathematics}. 2007.
\newblock Full reference to be find at
  http://www.math.uh.edu/~mg/reprints/abstracts/06ADGW.html.

\bibitem{Kopell2000}
N.~Kopell and G.B. Ermentrout.
\newblock Mechanisms of phase-locking and frequency conrol in pairs of coupled
  neural oscillators.
\newblock In {\em Kandbook of Dynamical Systems}. 2000.

\bibitem{Kopell2004}
Nancy Kopell and Bard Ermentrout.
\newblock Chemical and electrical synapses perform complementary roles in the
  synchronization of interneuronal networks.
\newblock {\em PNSA}, 101(43):15482--15487, october 2004.

\bibitem{Galan2005}
Roberto~F Galán, G.~Bard Ermentrout, and Nathaniel~N Urban.
\newblock Efficient estimation of phase-resetting curves in real neurons and
  its significance for neural-network modeling.
\newblock {\em Phys Rev Lett}, 94(15):158101, Apr 2005.

\bibitem{Leutgeb2005}
Jill~K. Leutgeb, Stefan Leutgeb, Alessandro Treves, Retsina Meyer, Carol~A.
  Barnes, Bruce~L. McNaughton, May-Britt Moser, and Edvard~I. Moser.
\newblock Progressive transformation of hippocampal neuronal representations in
  "morphed" environments.
\newblock {\em Neuron}, 48:345--358, october 2005.

\bibitem{Gerstner2002}
Wulfram Gerstner and Werner Kistler.
\newblock {\em Spiking Neuron Models}.
\newblock Cambridge University Press, 2002.

\bibitem{Thorpe2001}
S.~Thorpe, A.~Delorme, and R.~Van Rullen.
\newblock Spike-based strategies for rapid processing.
\newblock {\em Neural Netw}, 14(6-7):715–725, 2001.

\bibitem{Carnell2005}
Andrew Carnell and Daniel Richardson.
\newblock Parallel computation in spiking neural nets.
\newblock May 2005.

\bibitem{Jin2004}
Dezhe~Z. Jin.
\newblock Spiking neural network for recognizing spatiotemporal sequences of
  spikes.
\newblock {\em PHYSICAL REVIEW}, 69, 2004.

\bibitem{Aviel2003}
Y.~Aviel, C.~Mehring, M.~Abeles, and D.~Horn.
\newblock On embedding synfire chains in a balanced network.
\newblock {\em Neural Computation}, 15:1321--1340, 2003.

\bibitem{Toyoizumi2005}
Taro Toyoizumi, Jean-Pascal Pfister, Kazuyuki Aihara, and Wulfram Gerstner.
\newblock Generalized bienenstock-cooper-munro rule for spiking neurons that
  maximizes information transmission.
\newblock {\em PNAS}, 102(14):5239--5244, April 2005.

\end{thebibliography}

\end{document}